\documentclass[12pt]{article}
\usepackage{amssymb}
\usepackage{amsfonts}
\usepackage{epsfig}
\newcommand{\be}{\begin{eqnarray}}
\newcommand{\ee}{\end{eqnarray}}
\newcommand{\bez}{\begin{eqnarray*}}
\newcommand{\eez}{\end{eqnarray*}}
\renewcommand{\O}{\Omega}
\newcommand{\A}{{\cal A}}
\newcommand{\M}{{\cal M}}
\newcommand{\X}{{\cal X}}
\renewcommand{\O}{\Omega}
\renewcommand{\b}{\bar{b}}
\renewcommand{\a}{\tilde{a}}
\renewcommand{\d}{{\rm d}}
\newcommand{\pa}{\partial}
\newcommand{\bu}{\bullet}

\newcommand{\eq}[1]{(\ref{#1})}

\begin{document}

\thispagestyle{empty}
\renewcommand{\theequation}{\arabic{section}.\arabic{equation}}
\renewcommand{\thefootnote}{\fnsymbol{footnote}}

\vspace*{2.5cm}
\begin{center}

{\Large \bf Dynamical Evolution in Noncommutative \\ Discrete
Phase Space and the Derivation \\[2mm] of Classical Kinetic
Equations}

\vspace*{0.8cm}

{\bf A. Dimakis}\footnote{dimakis@aegean.gr}
 \\ Department of Mathematics, University of the Aegean \\
        GR-83200 Karlovasi, Samos, Greece \\

\vspace*{0.5cm}

       {\bf C. Tzanakis}\footnote{tzanakis@edc.uoc.gr}
        \\ Department of Education, University of Crete \\
        GR-74100 Rethymnon, Crete, Greece \\

\vspace*{1cm}

{\bf Abstract}
\end{center}

{\footnotesize \noindent
By considering a lattice model of
extended phase space, and using techniques of noncommutative
differential geometry, we are led to: (a) the conception of vector
fields as generators of motion {\em and} transition probability
distributions on the lattice; (b) the emergence of the time
direction  on the basis of the encoding of probabilities in the
lattice structure; (c) the general prescription for the
observables' evolution in analogy with classical dynamics. We show
that, in the limit of a continuous description, these results lead
to the time evolution of observables in terms of (the adjoint of)
generalized Fokker-Planck equations having: (1) a diffusion
coefficient given by the limit of the correlation matrix of the
lattice coordinates with respect to the probability distribution
associated with the generator of motion; (2) a drift term given by
the microscopic average of the dynamical equations in the present
context. These results are applied to 1D and 2D problems.
Specifically, we derive: (I) The equations of diffusion,
Smoluchowski and Fokker-Planck in velocity space, thus indicating
the way random walk models are incorporated in the present
context; (II) Kramers' equation, by further assuming that, motion
is deterministic  in coordinate space}.
\vspace{2cm}

\pagebreak \setcounter{page}{1}
\renewcommand{\thefootnote}{\arabic{footnote}}
\setcounter{footnote}{0}

\section{Introduction}
\setcounter{equation}{0}

In the last few years there has been an increasing interest in
exploring the possible relevance of noncommutative geometry (NCG)
in various areas of physics. Roughly speaking, the basic idea in
this context, is to try to develop an appropriate conceptual and
mathematical framework in which the fundamental object is no
longer a manifold (intuitively conceived as the generalization of
a geometrical surface), but rather an appropriate algebra
$\cal{A}$ (e.g.\ of $\mathbb{R}$ or $\mathbb{C}$ functions on it).
This is (partly) motivated by the well-known result that
commutative $C^*$-algebras correspond biuniquely to locally
compact topological spaces, hence that all relevant information
for the topological structure of such spaces is encoded in their
algebra of functions (e.g. \cite{Bratelli}).

From an algebraic point of view there are two possible lines of
approach, depending on the commutativity or noncommutativity  of
the algebras considered. Though the latter has received
considerable attention, especially in connection with the study of
quantum groups and quantum field theory
(\cite{ConLo1991,Con1994,Mad1995,Lan1997,Var1997} and references
therein), it is true that even if the algebra $\cal{A}$ is {\em
commutative}, the resulting structures are quite rich, allowing
for a considerable generalization of familiar geometric concepts,
while keeping a rather close contact with ordinary (commutative)
differential geometry. In particular, geometrical concepts and
methods may be developed on both ``continuous'' and discrete sets,
with interesting applications in such diverse fields as riemannian
geometry, gauge field theory, integrable dynamical systems,
stochastic calculus, symplectic mechanics and kinetic theory
(\cite{DM-H1992a}--\cite{DM-H1999}; for a review see
\cite{M-H1997}, \cite{DM-H1997}  and references there in). In this
connection, a basic tool is provided by the (great variety of)
differential calculi that can be defined on a commutative algebra
$\cal{A}$  with unit $1$, generalizing the familiar (exterior)
calculus of differential forms. More specifically, a differential
calculus (DC) on a commutative algebra  $\cal{A}$, is a graded
associative, in general {\em noncommutative} algebra \bez
 \O(\A) :=\bigoplus_{r\in \mathbb{N}_0}\O^r(\A)
  \eez
where $\O^0(\A)=\A$, $\O^r(\A)$ is an $\A$-bimodule, equipped with
an exterior derivative, i.e.\ a linear operator
\be
\d:\O\to\O,\qquad\qquad \d^2=0, \;\;\; \d(1)=0 \label{1.1} \ee
\be
\d(\omega\omega')=(\d\omega)\omega'+(-1)^r\omega\,\d\omega',\qquad
\omega\in\O^r(\A) \label{1.2} \ee Elements of $\O^r(\A)$ are
called (differential) $r$-forms and we assume that as an
$\A$-bimodule, $\O^r(\A)$ is generated by $\d(\O^{r-1}(\A))$.

A key concept here, allowing for interesting conceptual insights,
is the {\em commutative} and {\em associative} product of 1-forms
defined by\footnote{ From now on we write $\O^r$ instead of
$\O^r(\A)$ and the symbol ``$:=$'' indicates a defining relation
for the object that lies on the side of ``:''.}
\be
\omega \bullet \d f := [\omega, f] := \omega f-f \omega, \qquad f
\in \A, \, \omega \in \O^1 \label{1.3} \ee and extended bilinearly
in $\A$, so that
\be
(f \omega g) \bullet (f' \omega ' g') = ff' (\omega \bullet \omega
') gg', \quad f, f', g, g' \in \A, \; \omega , \omega ' \in \O^1
\label{1.4} \ee (for more details, see \cite{BDM-H1995}, \S\S2,
3). For the existence of this $\bullet$ product, commutativity of
$\A$ is essential.

In the usual DC, eq(\ref{1.3}) is zero so that $\bu$ is trivial.
In fact, by (\ref{1.3}), the Leibniz rule (\ref{1.2}) can be
written in $\O^1$ as \be \d (fg)= f \d g + g \d f + \d f \bu \d g
\label{1.4'} \ee Thus (\ref{1.4'}) stresses the nature of
noncommutative DC as a {\em deformation} of the usual DC. On the
other hand, it is formally identical with the generalized It\^{o}
differentiation rule in Stochastic Calculus (StC), $f,g$ being
semimartingales and the 3rd term in (\ref{1.4'}) is the so-called
bracket of the processes, related to their quadratic variation
(\cite{E1989}, appendix eq(4)). In fact, it can be shown that this
is not merely a formal analogy; more precisely, a noncommutative
product can be easily defined for semimartingales and their
differentials, so that the stochastic differentiation rule becomes
identical to (\ref{1.4'}) (\cite{DM-H1993} \S3). In this way one
can obtain results of StC by employing techniques of NCG (e.g. see
\cite{DM-H1993b} \S3 for the Ornstein-Uhlenbeck process)

Motivated by the above remarks, one may develop differential
geometry on the basis of a {\em minimal} deformation of the usual
DC, namely, a calculus in which
\be
\d f \bu \d g \bu \d h = [[\d h, g], f] =0 \label{1.5} \ee and
explore its relation to StC. Further motivation for this, stems
from the following remarks:

From a physical point of view, StC is a mathematical formalism on
the basis of which a precise meaning can be given to stochastic
dynamical models of physical systems whose time evolution cannot
be given in terms of deterministic flows in their phase space
(e.g.\ brownian motion). These are systems for which the concept
of a trajectory in phase space is not defined. Mathematically
speaking, their evolution cannot be described by semigroups of
Perron-Frobenius operators, since their states often evolve under
a 2nd order differential operator (differential operators can
generate Perron-Frobenius semigroups only if they are of the first
order, see e.g.\ \cite{LM1985}, ch.7). The appearance of such
operators is a basic feature of StC (via It\^{o}'s formula). In
this way, by appropriate generalizations of basic geometrical
concepts in the context of StC, a general mathematical framework
results for the description of such nondeterministic systems
(\cite{E1989}). On the other hand, 2nd order evolution equations
are at the heart of kinetic theory, for describing
time-irreversible evolution towards equilibrium; they are obtained
by employing some {\em approximation} scheme to the {\em
time-reversible} microscopic dynamics.

Taking account of the above discussion and starting with
(\ref{1.5}), we have formulated basic concepts of differential
geometry, tensor calculus and symplectic mechanics in a
noncommutative context and we have shown that hamiltonian
equations for observables are (adjoint to) generalized
Fokker-Planck (FP) equations like those encountered in It\^{o}'s
StC and the kinetic theory of open systems (\cite{DT1996},
\cite{DT1997}).

These are suggestive mathematical developments for the relevance
of NCG to StC and kinetic theory. However, from a conceptual point
of view, the physical meaning of noncommutativity has to be made
clearer, particularly in connection with the time evolution of
physical systems. On this issue, a basic intuitive idea is that
noncommutativity is due to the fact that differentials (i.e.\
1-forms) have a ``size'', so that different results are obtained
when functions are evaluated at the left end of the 1-form (its
multiplication by functions from the left) and at the right end of
it (multiplication from the right) - cf.\cite{DT1996}, \S6. On the
other hand, 2nd order FP equations result in the context of
kinetic theory (based either on microscopic dynamics or stochastic
models), by assuming that, although individual microscopic
interactions happen on a time-scale very short compared to the
(macroscopic) relaxation time of the system as a whole, yet this
microscopic time-scale is long enough to allow for many changes in
the {\em microscopic} configurations of the system (see e.g.
\cite{C1943} \S\S II.1, II.4, \cite{RedeL1978} ch.IX.4).
 Therefore, although eventually one
passes to the limit of a continuous description in terms of
macroscopic kinetic equations, the fine structure of the system on
the ``infinitesimal'' microscopic time-scale has already been
taken into account. This is a methodology widely used in
statistical mechanics and kinetic theory; to start from a {\em
discretized} picture of the system and subsequently pass to a {\em
continuous} macroscopic description in some appropriate limit
(e.g.\ thermodynamic limit, appropriate scaling of the physical
quantities, \cite{Bal1975} ch.3.3, \cite{Spo1980}, \S I,
\cite{Spo1991}, \cite{Li1969} ch.III).

On the basis of this, we may try to develop a {\em discrete}
analogue of the formalism in \cite{DT1996}, \cite{DT1997}, aiming
at the following: (a) To throw more light into the physical
meaning of noncommutativity of the DC; (b) to understand better
the similarity of the DC to StC and (c) to provide a method for
deriving irreversible evolution equations, which describe
nondeterministic motions and which can be given a probabilistic
interpretation, as {\em exact} rather than approximate results,
much in the spirit of \cite{DT1996}, \S1, and in contrast with the
general methodology underlying statistical mechanics (cf. the
discussion in \cite{DT1996} \S1).\footnote{Notice that, implicit
to the derivation of kinetic equations by approximating
microscopic dynamics, is the assumption that the higher the
approximation, the better are the resulting equations. However,
this usually leads to differential equations of order higher than
the second, in direct conflict with the fact that at least for
linear autonomous equations, it is known that only differential
operators of order at most 2 can generate semigroups of solutions
admitting a probabilistic interpretation, \cite{TGr1999}.} In this
connection, the techniques developed in \cite{DM-H1994},
\cite{DM-H1999} for NCG on discrete sets will be used, since, on
the one hand they allow for a clear geometric representation of
noncommutative DC and on the other hand, they are conceptually
closer to those used in the study of specific models in stochastic
dynamics or statistical mechanics.\footnote{Though formalisms
based on the ``continuum'', more often than not can be
 manipulated mathematically more easily than formalisms based
on
 discrete methods, the latter often provide deeper conceptual
insights.}

In view of the above, in the present paper we will show how simple
model kinetic equations can be derived in the context of DC on a
discrete set, having a three-fold aim:\\ (i) From the {\em
conceptual point of view}, to clarify the relevance of
noncommutative DC to kinetic theory and more generally, to the
description of physical systems in terms of irreversible evolution
equations;\\ (ii) since in kinetic theory and StC, probabilistic
concepts are central, to explain how such concepts could naturally
arise in the context of DC applied to systems conventionally
studied by kinetic theory and stochastic dynamics;\\ (iii) in
analogy with classical dynamics, to formulate a {\em general
prescription} for what dynamical evolution means in a formalism
based on NCG.

More specifically, the paper is organized as follows:\\ The basic
ideas in this section and the above general aims of the paper, are
already apparent in the 1D-model considered in section 2. This
fact suggests the line of approach in the subsequent sections. The
basic ideas are introduced systematically in section 3.
Specifically, (a) the extended phase space as a discrete set, in
particular a hypercubic oriented lattice, (b) vector fields  as
{\em generators of motion and probability distributions},
 (c) {\em the emergence of the time direction} on
the basis of the encoding of probabilities in the lattice
structure of the extended phase space, (d) the general
prescription of the {\em evolution equation for observables}, in
direct analogy with conventional dynamics. In section 4 we apply
the previous ideas and results in 1D-problems, thus getting the
incorporation of random walk models in our context and accordingly
showing how the diffusion, Smoluchowski and FP equation in
velocity space, result in the continuous limit. In section 5 the
approach of section 4 is generalized in $N$ dimensions, giving in
the continuous limit a generalized FP equation in which, (1) the
diffusion matrix is nonnegative-definite and is the limit of the
correlation matrix of the lattice, with respect to the probability
distribution associated with the vector field\  which is the
generator of motion; (2) the drift terms are specified by the
dynamics of the particular problem under investigation. Some
comments on the form and properties of the transformations in
phase space, are given in section 6. In section 7 the results of
section 5 are applied to the 2D-case. As an example, Kramers'
equation is derived by assuming Newtonian equations of motion with
friction, and requiring that in the continuous limit, motion in
{\em coordinate} space is deterministic (i.e.\ trajectories exist
there). In section 8 the nature and mathematical properties of the
limiting procedure used in this paper and of other such
procedures, are examined, showing that only the limit used here is
well defined for {\em all} possible (1st order) differential
structures on the lattice. Moreover, if the conceptual framework
of \S3 is employed, then
 we show that the only differential operators which can be generators of evolution of observables are
 at most of order 2. Finally, in section 9 we summarize our
results and discuss the main ideas involved, as well as possible
further elaborations in the present context.

\section{A One-Dimensional Model: Motivation For Further Developments}
\setcounter{equation}{0}

As already mentioned in \S1, there is a formal similarity of the
Leibniz rule in noncommutative DC, eq(\ref{1.4'}), and stochastic
differentiation. Actually, in \cite{DM-H1993}, it was shown that
It\^{o}'s stochastic differentiation can be made an exterior
derivative in noncommutative DC. In this section we explore
further this issue by means of a simple 1D {\em discrete} model,
following the rationale of \S1. Specifically, we consider a point
moving in one dimension $x$, so that motion can be described in
$\mathbb{R}^2$ with local coordinates $(t,x)$, equipped with a DC
having the following commutation relations ($t$ is the time; for
the choice of the signs see \S3.4)
\be
\d t \bu \d t = -b\, \d t, \quad \d x \bu \d x = - h\, \d t, \quad
\d t \bu \d x = -b\, \d x, \qquad h, b >0 \label{2.1} \ee and we
assume that $\{ \d t, \d x \}$ form a basis of 1-forms.
Associativity of $\bu$ implies that
\be
\d x \bu \d x \bu \d x = a^2 \d x, \qquad\qquad a^2:= h b
\label{2.2} \ee so that in the limit $a, b \rightarrow 0, \; a^2/b
\rightarrow h =\mathnormal{constant}$, we recover in one
dimension, the DC of \cite{DM-H1993} \S5, \cite{DT1996} \S4, in
close analogy with the It\^{o} StC and with commutation relations
\be
\d t \bu \d t =\d t \bu \d x =0, \qquad\qquad  \d x \bu \d x = - h
\,\d t \label{2.3} \ee It is a special case of the DC defined by
(\ref{1.5}). With the transformation
\be
u={1 \over 2} \left( {x \over a}- {t \over b} \right), \; v=-{1
\over 2} \left( {x \over a}+ {t \over b}\right) \;\;
\Leftrightarrow \;\; t=-b(u+v), \; x=a(u-v) \label{2.4} \ee
(\ref{2.1}) becomes
\be
\d u \bu \d u = \d u, \qquad \d u \bu \d v = 0, \qquad \d v \bu \d
v = \d v \label{2.5} \ee which is the DC on an oriented square
lattice. Hence, from (\ref{1.3}) we easily get
\be
\d f= \left( f(u+1,v)-f(u,v) \right) \d u + \left( f(u,v+1)-f(u,v)
\right)\d v \label{2.6} \ee so that a direct calculation using
(\ref{2.6}) gives
\be
\d f(t,x) =\,\bar{\pa}_x f(t-b,x) \d x + \,\left(\pa_{-t}f(t,x)-{h
\over 2}\Delta_x f(t-b,x)\right)\d t \label{2.7} \ee where
\be
(\bar{\pa}_x f)(t,x) := {f(t,x+a)-f(t,x-a)\over 2 a}, \label{2.8}
\ee
\be
(\pa_{-t}f)(t,x) :={f(t,x)-f(t-b,x)\over b}  \label{2.9} \ee
\be
(\Delta_x f)(t,x):= {f(t,x+a)+f(t,x-a)-2f(t,x)\over a^2}
\label{2.10} \ee In the limit $a, b \rightarrow 0, \, a^2/b
\rightarrow h =\mathnormal{constant}$, (\ref{2.7}) becomes
\be
\d f(t,x) =\,\pa_x f(t,x) \d x + \,\left(\pa_tf(t,x)-{h\over
2}\pa^2_x f(t,x)\right)\d t \label{2.11} \ee recovering in 1D the
results of previous works (\cite{DT1996} eq(4.13), \cite{DM-H1993}
eq(5.3), \cite{DM-H1993b} eq(5.1)) and essentially It\^{o}'s
formula for the differential of a function of a Wiener process
with variance $h$.

In order to explore this similarity further, we consider the
(free) motion of a point in this context, by introducing the
concept of a differentiable motion in analogy with the ordinary
DC. To this end, let $(\mathbb{R}, s, \d ')$ be a discrete DC on
$\mathbb{R}$ ($s$ being a coordinate function on $\mathbb{R}$
parameterizing this motion), defined by
\be
\d ' s \bu \d ' s =-b\, \d ' s \label{2.12} \ee Then, we define a
differentiable motion to be a mapping
\be
\gamma : \mathbb{R} \to \mathbb{R}^2\,,  \qquad\qquad \gamma (s) =
(t(s), x(s))
 \label{2.13}
\ee inducing a ``pull-back'' homomorphism $\gamma_\ast$ on
$\O^1(\mathbb{R}^2, \d)$ to $\O^1(\mathbb{R}, \d ')$
with\footnote{More generally, a mapping  $ \phi : (\O ' (\A), \d '
) \to (\O(\A), \d )$ between two DC, is called differentiable, if
it is a homomorphism of $\O'(\A)$ to  $\O (\A)$  and $ \phi \circ
\d ' = \d \circ \phi$ (see \cite{DM-HV1995}, \S4).}
\be
\gamma_\ast (x)=x(s), \quad \gamma_\ast(t)=t(s), \quad \gamma_\ast
\circ \d = \d ' \circ \gamma_\ast \label{2.14} \ee Eq(\ref{2.13})
implies
\be
\gamma_\ast(\d x) = {x(s)-x(s-b) \over b}\, \d 's =: \dot{x}(s) \d
's, \quad \gamma_\ast(\d t) = {t(s)-t(s-b) \over b}\, \d 's =:
\dot{t}(s)\d 's \label{2.15} \ee Applying $\gamma_\ast$ to the
commutation relations (\ref{2.1}), (\ref{2.2}), we get the
following consistency relations
\be
b \,\dot{x}^2(s) = h\,\dot{t}(s), \quad b\, \dot{x}(s)
\dot{t}(s)=b\, \dot{x}(s),
 \quad b\, \dot{t}^2(s)=b\, \dot{t}(s) \label{2.16}
\ee Therefore, with $\dot{x}(s)\neq 0, \, \dot{t}(s)\neq 0$ to
avoid trivial cases, we get $ \dot{t}=1, \; \dot{x} = \pm a/b$,
hence
\be
{x(s) - x(s-b) \over b} = \pm {a \over b} \label{2.17} \ee

The following remarks can be made here:
\smallskip

\noindent (i) Motion consists of jumps $\pm a$ in discrete time
intervals $b$, i.e (\ref{2.17}) gives the discrete velocity of the
point. In the continuous limit which leads to It\^{o}'s StC,
(\ref{2.17}) becomes infinite, strongly suggesting the picture of
brownian motion as a random walk with equal left and right
probabilities of jumps $\pm a$ at time intervals $b$ and diffusion
constant $a^2/b$ (see e.g.\ \cite{C1943} eq(17), \cite{Kac1956},
eq(7)).
\smallskip

\noindent (ii) The DC defined by (\ref{2.1}) is up to a coordinate
transformation the DC on an oriented square lattice (cf.\
(\ref{2.4})).
\smallskip

\noindent If the model is to be taken seriously, then these
remarks raise the following issues:
\begin{itemize}
\item How can probabilistic concepts be introduced, given that no such
concepts appeared above? In other words, is there some deeper
reason for the formal similarity of the model with random walk,
which may lead to the introduction of probabilities at a
fundamental level?
\item The above model corresponds to {\em 1D-free} motion. How can
it be generalized to {\em higher} dimensions and/or {\em other
models}? That is, what is the general prescription of time
evolution in the context of noncommutative DC?
\end{itemize}
These questions correspond to the general aims of the paper,
expressed in \S1 and are discussed in the next section.

\section{Basic Ideas And General Formalism}
\setcounter{equation}{0}

In \S1 we noticed that in statistical mechanics and kinetic
theory, one often starts from a {\em discrete} picture of the
system under consideration, either for methodological reasons, or
because at the microscopic level these systems are in some
fundamental sense discrete (e.g.\ composed by a number of
particles). This idea also appears in other areas, like lattice
field theory (e.g.\ \cite{Roe1994}, ch.8), or certain approaches
to quantum gravity (e.g.\ \cite{Bomb1987,Sork1991}, see also
\cite{Loll1998}, \cite{Gibbs1996} for recent reviews, and
references in \cite{DM-HV1995} \S1). Therefore, motivated by the
discussion in \S\S1 and 2, we start with a discrete (finite or
denumerable) set $\M$, which we call the ``phase space'' and
eventually pass to the special case of an oriented hypercubic
lattice $\mathbb{Z}^{N+1}$. We will use the concepts and
techniques developed in \cite{DM-H1994,DM-HV1995,DM-H1999}, a
summary of which is given below. At this point we stress that, as
explained in \S3.1, starting with a discrete picture has
far-reaching consequences, since any differential structure on
$\M$ is {\em necessarily noncommutative}, i.e.\ discretization
implies noncommutativity of the DC on $\M$. Thus, the need for, or
necessity of, a discrete description of physical objects, can be
seen as a basic motivation (or explanation!) for introducing
noncommutative differential structures in the study of physical
systems.

\subsection{Differential calculi on a discrete set: A summary}

Let $\A$ be the algebra of $\mathbb{C}$-valued functions on $\M$,
with the usual algebraic operations. Then,
\be
f= \sum_{i \in \M} f(i) e_i =: \sum_{i \in \M} f_i e_i, \quad
\forall f \in \A \ \qquad \qquad e_i(j) := \delta_{ij}\,,\quad e_i
\in \A \label{3.1} \ee
\be
e_i e_j = \delta_{ij} e_i \, , \qquad\qquad \sum_i e_i =1
\label{3.3} \ee We introduce a DC on $\A$ as in \S1. By defining
\be
e_{ij} := \left \{ \begin{array}{ccc} e_i \d e_j & & i \neq j\\ 0
& & i=j  \end{array} \right.  \label{3.4} \ee so that
\be
f\, e_{ij} = f_i \,e_{ij}, \qquad\qquad  e_{ij}\, f= f_j\, e_{ij}
\label{3.5} \ee we get $e_i \d e_i = - \sum_{j \neq i} e_{ij}$ and
consequently
\be
\d e_i = \sum_j (e_{ji}-e_{ij}) \label{3.6} \ee On the basis of
this we can prove that $\{e_i\}$ is linearly independent and a
basis of $\O^1$ taken as a vector space over $\mathbb{C}$. In
particular
\be
\d f= \sum_{i,j} (f_j -f_i) e_{ij} \, , \quad \d f \bu e_{ij} =
(f_j -f_i) e_{ij} \, , \quad e_{ij} \bu e_{kl} = \delta_{ik}
\delta_{jl} e_{ij} \, , \qquad \forall f \in \A \label{3.7} \ee It
can be proved (\cite{DM-H1994} \S II) that the $e_{ij}$ induce a
basis on $\O^r, \ \ r \geq 2$ over $\mathbb{C}$ via
\be
\{ e_{i_1\cdots i_r} |\; e_{i_1\cdots i_r}:= e_{i_1i_2}
e_{i_2i_3}\cdots e_{i_{r-1}i_r} \} \label{3.8} \ee Therefore,
there is a simple way to obtain any particular DC on $\A$ simply
by imposing relations among the $e_{ij}$, which turn to be
equivalent to putting some of the $e_{ij}$ equal to $0$. This
implies additional form relations for $r \geq 2$. Actually, the DC
defined by (\ref{3.4}), (\ref{3.8}) is the largest one, called the
{\em universal} DC on $\A$. At this point it is conceptually
suggestive to notice that $e_{ij}$ with $i \neq j$ may be
represented as an arrow from $i$ to $j$, $i \rightarrowtail j$.
Then, universality means that all pairs of points of $\M$ are
connected by two antiparralel arrows and any DC on $\A$ is
obtained by simply discarding  some arrows from this complete
di-graph (viz. directed graph). In this representation,
(\ref{3.6}) gives
\bez
\d e_i= \mbox{(sum of incoming arrows at $i$)$
- $ (sum of outgoing arrows at $i$)}
\eez
and similar
interpretations hold for (\ref{3.8}). Moreover, (\ref{3.5}) shows
that left (right) multiplication of 1-forms $e_{ij}$ by functions,
implies that functions are evaluated at the starting (end) point
of the corresponding arrow. This proves that {\em any} nontrivial
DC on a discrete set $\M$ is {\em necessarily} noncommutative,
since at least one $e_{ij}$ is nonzero so that $\d f \bu e_{ij}$
is not identically zero. Actually, this quantity is (proportional
to) the {\em change} of $f$ along the arrow $i \rightarrowtail j$.
These remarks are important for what follows in this section.

\subsection{Vector fields on a discrete set $\M$}

Vector fields $X$ on $\A$ are elements of the $\A$-bimodule
$\cal{X}(\A)$ dual to $\O^1$, defined by
\be
\langle f \omega , g X \cdot h \rangle := f g\langle \omega h \, ,
X \rangle  \, ,  \qquad f, g, h \in  \A, \;\omega \in \O^1
\label{3.9} \ee where $\langle\,,\rangle$ denotes duality
contraction with $\O^1$ considered as a {\em left} $\A$-module.

By (\ref{3.9}), vector fields\  act as operators on $\A$
\be
X(f) := \langle \d f , X \rangle  \label{3.10} \ee hence by
(\ref{3.9}) and the Leibniz rule
\be
(fX) (g) = f X(g) \,, \quad\qquad (X\cdot f)(g) = X(f g) - g X(f)
\label{3.11} \ee We define elements $\{ \pa_{ij} \} \in \X(\A)$,
``dual'' to $\{e_{ij}\}$, by
\be
\langle e_{ij}, \pa_{kl} \rangle = \delta_{ik} \delta_{jl} e_i
\label{3.12} \ee (putting $\pa_{kk} = 0$). A direct calculation,
using (\ref{3.12}), (\ref{3.5}), gives
\be
e_k  \pa_{ij} = \delta_{ik}  \pa_{ij} \, , \qquad \qquad \pa_{ij}
\cdot e_k = \delta_{jk} \pa_{ij} \label{3.13} \ee so that
\be
\pa_{ij} \cdot f = f(j)  \pa_{ij} \, , \quad f  \pa_{ij} =  f(i)
\pa_{ij} \, , \qquad \pa_{ij}(f) = (f(j)-f(i)) e_i \label {3.14}
\ee Hence for any $X \in \X$
\be
X= \sum_{i,j} X^{ij} \pa_{ij} \, , \qquad X(f) = \sum_{i,j}
(f_j-f_i) X^{ij} e_i \, , \qquad  X^{ij} \in \mathbb{C}
\label{3.15} \ee and we may write $X$ as an ordinary difference
operator $X=\sum_{i,j} X^{ij} \pa_{ij} =: \sum_\alpha X^\alpha
\pa_\alpha$, where $\alpha$ is an index for the arrows of the
di-graph. The following proposition plays an important role in
this paper and its proof is straightforward:
\medskip

\noindent {\bf Proposition 3.1} {\em Let $I$ be the identity
mapping on $\A$\/. Then $\phi: \A \to \A$ is an  endomorphism of
$\A$, if and only if
\be
X:= \phi - I \label{3.16} \ee satisfies
\be
X(fg) = g\,X(f) + f\,X(g) + X(f)\, X(g) \label{3.17} \ee } We
remark here that $X$ in general is not a vector field\  on $\M$.
Nevertheless eq(\ref{3.14}) readily implies that $\pa_{ij}$
(equivalently, $\pa_\alpha$), satisfy (\ref{3.17}). Using this, we
easily prove:
\medskip

\noindent {\bf Proposition 3.2} {\em A vector field\  $X \in \X (
\A )$ generates an endomorphism $\phi$ via (\ref{3.16}), if and
only if
\be
X = \sum_\alpha X^\alpha \pa_\alpha \;, \qquad  \mbox{with} \qquad
X^\alpha X^\beta - \delta^{\alpha\beta} X^\alpha =0 \label{3.18}
\ee } Eq(\ref{3.18}) simply means that at each point of $\M$, {\em
at most one} of the $X^\alpha$ is nonzero and equal to 1. By
(\ref{3.17}), (\ref{3.18}), $X$ is the generator of a discrete
semigroup of endomorphisms of $\A$, $\{ \phi^n = (I+X)^n, \;  n
\in \mathbb{N} \}$.

As it is shown in appendix A.1, $\phi$ is an automorphism, if and
only if it induces a 1-1 and onto mapping $\Phi \!: \M \to \M$. In
this case, $\phi$ is essentially the Koopman operator associated
with $\Phi$ and maps the basis $\{ e_i| \; i \in \M \}$ onto
itself (see e.g. \cite{LM1985} \S3.3). Therefore, the orbits of
the semigroup generated by $X$ in (\ref{3.18}) are determined by
the trajectories of points of $\M$ under $\Phi$ (i.e.\ the flow
defined by $\Phi$), which may be thought as lying along arrows of
the di-graph corresponding to the DC on $\M$ (notice that by the
remark above, at each point of $\M$ at most one term in
(\ref{3.18}i) is nonzero, and by (\ref{3.14}) each $\pa_{ij}$
gives the change of functions along the arrow $i \rightarrowtail
j$). Thus, vector fields\ generating automorphisms of $\A$, define
flows on $\M$ along arrows of the di-graph that map points of $\M$
to points of $\M$. In this picture, stemming from the discrete
character of $\M$, motion along a trajectory defined by a vector
field\ $X$, is {\em constrained} to be along arrows of the
di-graph, so that at each point $i$ of $\M$ (i.e.\ vertex of the
di-graph), motion along any arrow is either {\em certain} or {\em
impossible}, depending on $X$.

The above discussion raises naturally the question {\em what kind
of ``motion'' is described by an arbitrary vector field, i.e.\ one
not necessarily satisfying (\ref{3.18})?} Evidently, $\{ e_i \}$
(equivalently, $\M$) is no longer mapped onto itself, but to a set
of nontrivial linear combinations of the $e_i$'s, that is, to a
``superposition of points'' (cf.\cite{DM-H1992a}, \S3). This
reminds the situation in statistical mechanics; non-unitary
semigroups of operators on the state space, transform  pure states
into mixtures in the quantum case and  $\delta$-distributions into
more general ones in the classical case.

On the basis of this analogy and the discussion in the previous
paragraph, we shall interpret {\em a vector field\ $X$ as the
generator of evolution on $\M$, having at each point a particular
{well-defined} probability associated with each arrow that emerges
from that point}\/. More precisely, in this interpretation, {\em
the component $X^{ij}$ of $X$ gives the transition probability for
the ``infinitesimal'' motion from $i$ to $j$}\/.

In this picture, vector fields\ $X$ acquire a {\em double} role;
as {\em generators} of evolution and as {\em states} giving the
transition probabilities for ``infinitesimal'' changes. This
double role is reflected in the notation of (\ref{3.10}) so that
$\langle \d f , X \rangle = X(f)$ is the expectation (average)
value for the ``infinitesimal'' change of the observable $f$ along
$X$ (cf.\ (\ref{3.15})). It is a key idea following from adopting
a discrete structure as our starting point. More precisely, though
both in the discrete and the usual continuous case, vector fields\
are linear combinations of 1st order  difference and differential
operators, respectively, it is {\em only in the continuous case}
that {\em any} such {\em combination generates automorphisms} of
the algebra of functions, or equivalently, {\em flows} of
trajectories (i.e.\ what we call here a {\em deterministic}
evolution, or motion). This is due to the fact that, in contrast
to the continuum, nontrivial linear combinations of displacements
do not give a permissible displacement on the di-graph. In the
rest of this section we shall pursue further this idea.

\subsection{The oriented hypercubic lattice and vector fields on it}

To be more specific and for applications in subsequent sections,
we shall henceforth take $\M$ to be an oriented hypercubic
lattice; that is, $\M = \mathbb{Z}^N$ with elements $\vec{k}\in\M$
and
\be
e_{\vec{k} \vec{l}} \neq 0 \quad \Leftrightarrow \quad \vec{l}=
\vec{k}+ \hat{\mu} ,\qquad\quad \hat{\mu} := (\delta_\mu^\nu)
\label{3.19} \ee $\{  \hat{\mu} \}$ being the canonical basis of
$\mathbb{R}^N$ (for details see \cite{DM-H1994}). Then, by
defining
\be
u^\mu := \sum_{\vec{k}} k^\mu e_{\vec{k}} \label{3.20} \ee and
writing $u:=(u^\mu)$, we get for $f \in \A$
\be
f = \sum _{\vec{k}} f(\vec{k}) e_{\vec{k}} = f(u) \label{3.21} \ee
so that from (\ref{3.7}), (\ref{3.15}) we get
\be
\d u^\mu = \sum_{\vec{k}} e_{\vec{k}, \vec{k} + \hat{\mu}}
\label{3.21'} \ee
\be
\d f &=& \sum_{\vec{k}, \mu} (f(\vec{k}+ \hat{\mu}) -
f(\vec{k}))e_{\vec{k}, \vec{k} + \hat{\mu}}
 = \sum_\mu (f(u+\hat{\mu})-f(u)) \d u^\mu\nonumber\\
& =: & \sum_\mu (\pa_{+u^\mu} f) \d u^\mu \label{3.22} \ee
\be
\pa_{+u^\mu} =& \sum_{\vec{k}} \pa_{\vec{k},\vec{k}+\hat{\mu}}
\label{3.23} \ee where we use the same symbol for $f$ as a
function of $\vec{k}$ and of $u$, to avoid a cumbersome notation.

Clearly, $\{ \d u^\mu \} $ is a basis for 1-forms with dual basis
for vector fields\  $\{ \pa_{+u^\mu} \}$
\be
X= \sum_{\vec{k}, \mu} X^{\vec{k},\vec{k}+\hat{\mu} } \,
\pa_{\vec{k},\vec{k}+\hat{\mu}} =: \sum_\mu P^\mu \pa_{+u^\mu}\,,
\qquad P^\mu = \sum_{\vec{k}} X^{\vec{k},\vec{k}+\hat{\mu}}
e_{\vec{k}} \label{3.24} \ee
\be
\langle \d u^\mu , \pa_{+u^\nu} \rangle = \delta^\mu_\nu
,\qquad\qquad \d u^\mu \bu \d u^\nu = \delta^{\mu \nu} \d u^\mu
\label{3.25} \ee the last relation following from (\ref{3.7}).

The interpretation of vector fields\ as states giving transition
probabilities, discussed in the previous subsection is manifested
in (\ref{3.24}); $X^{\vec{k}, \vec{k}+\hat{\mu}}$ is the
transition probability at a point $\vec{k}$ in the direction of
the axis $u^\mu$, so that $P^\mu$ is the distribution of this
probability on $\M$. Therefore, from now on we consider vector
fields\ $X$ such that
\be
P^\mu \geq 0 \; , \qquad \sum_\mu P^\mu =1 \label{3.26} \ee Thus,
in view of (\ref{3.22}), (\ref{3.25}), $\langle \d f,X \rangle$
gives at each point of $\M$, {\em the expectation of $\d f$ (the
``infinitesimal'' change of $f$), along $X$, with respect to the
probability distribution determined by $X$}\/. To see that this
interpretation is consistent and for later use as well, we notice
the following: The l.h.s.\ of (\ref{3.18}ii) is rewritten for an
arbitrary $X=\sum_\mu P^\mu \pa_{+u^\mu}$ as

\be
P^{\mu \nu} := \delta^{\mu \nu} P^\mu -  P^\mu  P^\nu = \langle \d
u^\mu \bu \d u^\nu , X \rangle - \langle \d u^\mu , X \rangle
\langle \d u^\nu , X \rangle , \label{3.27} \ee where \eq{3.25}
has been used. $(P^{\mu\nu})$ is the correlation matrix of the
$N$-dimensional random variable $(\d u^\mu)$ having a probability
distribution (determined by) $X$, where $\bu$ is the natural
(intrinsic) product of 1-forms. $P^{\mu \nu}$ vanishes if and only
if $X$ generates a flow of trajectories in $\M$. Actually, at
$\vec{k} \in \M$, $P^{\mu \nu}(\vec{k})$ is the correlation matrix
of the following random variables \bez
 I_\mu (\vec{k}) = \left \{ \begin{array}{cl}
1 &\mbox{if change at $\vec{k}$ is along the $u^\mu$ axis}\\ 0 &
\mbox{otherwise}  \end{array} \right. \eez with probabilities
$P^\mu(\vec{k})$ (cf.\ \cite{Ross1994}, \S3.3, p.331).

The following proposition has an interesting probabilistic
interpretation to be used later on
\medskip

\noindent {\bf Proposition 3.3} {\em  For $\omega= \sum_\mu s_\mu
\d u^\mu \; \in \O^1$, write $s^t =(s_\mu)$ (row vector). Then
\be
\langle \omega \bu \omega , X \rangle - (\langle \omega , X
\rangle)^2 = s^t \mathbb{P} s \label{3.28} \ee with $\mathbb{P} =
(P^{\mu \nu})$. Moreover, for
\be
\rho := \sum_\mu \d u^\mu
=\sum_{\vec{k},\mu}e_{\vec{k},\vec{k}+\hat{\mu}} \label{3.29} \ee
we have
\be
\rho \bu \omega = \omega, \qquad\qquad \forall \; \omega \in \O^1
\ee
\be
\langle \rho \bu \rho , X \rangle - (\langle \rho , X \rangle)^2
=0 \, , \quad\langle \rho \bu \omega , X \rangle - \langle \rho ,
X \rangle \langle \omega , X  \rangle =0, \qquad \forall \; \omega
\in \O^1 \label{3.31} \ee
\be
P^{\mu \nu}= \langle \left(\d u^\mu - \langle \d u^\mu , X \rangle
\rho \right)
 \bu \left(\d u^\nu - \langle \d u^\nu , X \rangle \rho \right), X  \rangle
 \label{3.31'}\ee
} The proof is immediate, but we stress the fact that, for
(\ref{3.31}) it is essential that $\sum_\mu P^\mu =1$.

Eq(\ref{3.31'}) is an equivalent expression of (\ref{3.27}),
obtained by using the unit $\rho$ with respect to the natural
multiplication $\bu$ of 1-forms. On the other hand, in the present
conceptual framework,
 (\ref{3.28}) gives the
variance of the random variable $(s_\mu)$ (and for that matter, of
$\omega$) having probability distribution $X$. Thus (\ref{3.31})
says that the unit of $(\O,\bu)$ has {\em zero variance} and is
{\em uncorrelated with all} $\omega \in \O^1$. Thus, {\em as a
random variable, $\rho$ is constant}, a fact that will be used in
the next subsection. Algebraically, (\ref{3.31}ii) says that
$\mathbb{P}$ has {\em always} a zero eigenvalue and a
corresponding 1D-eigenspace $\{ \lambda (1,1, \ldots , 1) \, ,
\lambda \in \mathbb{R} \}$. It is easily seen that this is the
only generic eigenspace of zero, in the sense that it is the {\em
only } one existing at {\em all} points of $\M$.

\subsection{The concept of time}

As already mentioned at the beginning of this section, we consider
$\M$ to be the {\em extended} phase space, in the sense that it
contains the ``time axis''. Adopting a newtonian picture we
require:

\noindent (a) For every evolution (e.g.\ motion) in $\M$, a {\em
change of time} is required;
\smallskip

\noindent (b) Time flows with certainty, that is, there is {\em
 always} a change in
the time axis.
\smallskip

\noindent Then, if any evolution on $\M$ is determined by some
vector field\ $X$, then (a) means that the time change is given by
1-forms $\tau$ such that $\langle \tau , \pa_{+u^\mu} \rangle \neq
0$ at all points of $\M$. Moreover, in view of the discussion
following proposition 3.3, (b) implies that
\be
\tau = \-b \rho = \d (-b \sum_\mu u^\mu) =: \d t \, , \quad
\langle \d t , X \rangle = -b \,, \quad \d t \bu \d t = -b \d t \,
, \qquad b \in \mathbb{R}^+ \label{3.32} \ee so that (a) is also
satisfied.

\noindent (i) Since $u^\mu \,, P^\mu$ are dimensionless, $b$ is
put for dimensional reasons.
\smallskip

\noindent (ii) By the interpretation of vector fields\ $X$ both as
generators of evolution and states, $X(f)=0$ should describe the
evolution of observables in the {\em extended} phase space. Then,
as explained in appendix A.2, $X(t)$ should be negative, in
analogy with classical dynamics, hence the choice of the $(-)$
sign in (\ref{3.32}).

\subsection{General prescription of dynamical evolution}

In summary, the approach described in this section leads to a
general prescription for time evolution in $\M$:\begin{itemize}
\item $\M$ is taken to be the extended phase space (cf.\ \S3.4).
\item Via (\ref{3.10}), vector fields\ $X$ are seen both as generators of
evolution and as states giving the transition probabilities for
``infinitesimal'' changes (cf.\ \S\S3.2, 3.3).
\item Time change appears in steps of duration $b$.
\end{itemize}
Therefore, elements $f$ of $\A$ are {\em observables}, whose
``infinitesimal'' change has an expectation value with respect to
$X$, $\langle \d f , X \rangle = X(f)$. But then, by remark (ii)
in \S3.4 and appendix A.2, $-X(f)/b$ gives the {\em rate of change
of $f$ along $X$ in extended phase space}, hence, in analogy with
classical dynamics, we take as the general dynamical evolution
equation
\be
\mbox{evolution equation for observable $f$:}\hspace{2cm} -{X(f)
\over b} = 0 \hspace{4.5cm} \label{3.33} \ee which becomes with
the aid of (\ref{3.22}), (\ref{3.24})
\be
-{1 \over b} \sum_\mu P^\mu(u) (f(u+\hat{\mu}) - f(u)) = 0
\label{3.34} \ee Eq(\ref{3.33}) is a general prescription,
independent of the choice of $\M$ as an oriented hypercubic
lattice. Nevertheless, it is our starting point in the
applications in the next sections. As we shall see there, in the
continuous limit,
 (\ref{3.34}) reduces in particular cases to the adjoint of well
known kinetic equations, i.e.\ to evolution equations for
observables.

\section{One Dimensional Problems}
\setcounter{equation}{0}

It is quite straightforward to apply the general ideas and results
of \S3 to specific cases. To keep technical details to a minimum,
we consider in this section as an illustration, the derivation of
well known 1D-model evolution equations and leave the general
treatment to the next sections.

Motivated by \S3.4 and the model of \S2, we consider a coordinate
system $(t,x)$ in the notation of \S2. Although the transformation
from the lattice axes to $(t,x)$ can be any invertible
transformation, we postpone this till the next section and use
``light-cone'' coordinates defined by (\ref{2.4}). Then, for $f
\in \A$, $\d f$ is given by (\ref{2.7}). On the other hand,
(\ref{3.24}) is
\be
X&=& p \pa_{+u} +q \pa_{+v} = \langle \d u , X \rangle \pa_{+u} +
\langle \d v , X \rangle \pa_{+v}  \nonumber \\ &=& \left(
\sum_{\vec{k}} X^{\vec{k},\vec{k}+\hat{1}} e_{\vec{k}} \right)
\pa_{+u}
 + \left( \sum_{\vec{k}} X^{\vec{k},\vec{k}+\hat{2}}
e_{\vec{k}} \right)\pa_{+v} , \label{4.1} \ee where $p,q\geq 0 $
and $p+q=1$. Using (\ref{2.7}) in (\ref{3.33}) and that $p+q=1$,
we readily obtain the evolution equation
\be
-{X(f) \over b} & =& \left(\pa_{-t}f(t,x)-{h\over 2}\Delta_x
f(t-b,x)\right) {\langle \d t , X \rangle \over -b} + \bar{\pa}_x
f(t-b,x){\langle \d x , X \rangle \over -b} \nonumber \\ &=&
\left(\pa_{-t}f(t,x)-{h \over 2}\Delta_x f(t-b,x)\right) - {a
\over b} (p-q) \bar{\pa}_x f(t-b,x)=0 \label{4.2} \ee In the
derivation of (\ref{4.2}), we remark that the following features
are present in any number of dimensions (see e.g.\ \S5 and
appendix B):\\ (a) Because of (\ref{3.26}), $\langle \d t , X
\rangle = -b$.\\ (b) 2nd order difference operators appear only in
the coefficient of $\d t$.\\ (c) the term involving $\langle \d x
, X \rangle = X(x)$ has the form of a drift term.\\ More
precisely, $\langle \d x , X \rangle$ has been interpreted in
\S3.5 as the ``infinitesimal'' change of $x$ in one step of time
$b$, hence it is natural to {\em assume} that it is {\em
proportional} to $b$. This assumption can be justified on the
basis of the general approach described in \S5 (see (\ref{5.*}),
the derivation of (\ref{5.12}) and appendix A.3). Since we
eventually pass to  the continuous limit considered in \S2, namely
$ a,\,b \to 0, \; \; a^2/b \to h=\mathnormal{constant}$, we have
\be
\langle \d x , X \rangle = a(p-q) = o(b)  \qquad \Rightarrow
\qquad p-q = o({a \over h}) = o(a) \label{4.3} \ee and therefore
\be
{\langle \d x , X \rangle  \over b}= {a \over b} (p-q) \to R \
\label{4.4} \ee say, when $a,\, b \to 0$, $a^2/b \rightarrow h$,
so that in this limit, (\ref{4.2}) becomes
\be
\pa_t f- R \pa_x f - {h \over 2} \pa_x^2 f  =0 \label{4.5} \ee To
see the physical meaning of this evolution equation, we go back to
(\ref{3.24}) and consider two special cases.
\medskip

\noindent {\bf 1.}
$X^{\vec{k},\vec{k}+\hat{1}},X^{\vec{k},\vec{k}+\hat{2}}$ are
constant:\\ Then from (\ref{4.1}), (\ref{4.3})
\be
X^{\vec{k},\vec{k}+\hat{1}}=p=: {1 \over 2} - \gamma a \, ,\quad
X^{\vec{k},\vec{k}+\hat{2}}=q=: {1 \over 2} + \gamma a \, , \qquad
\gamma = \mathnormal{constant} \label{4.6} \ee As it is evident
from the figure, $p,\,q$ represent the transition probability for
right and left motion respectively, along the $x$-axis. This is a
random walk model in 1D-position space and in view of (\ref{4.2}),
the evolution equation in the continuous limit, is
\be
\pa_t f +2\gamma h \pa_x f - {h\over 2} \pa_x^2 f  =0 \label{4.7}
\ee

\begin{figure}
\centering
\includegraphics{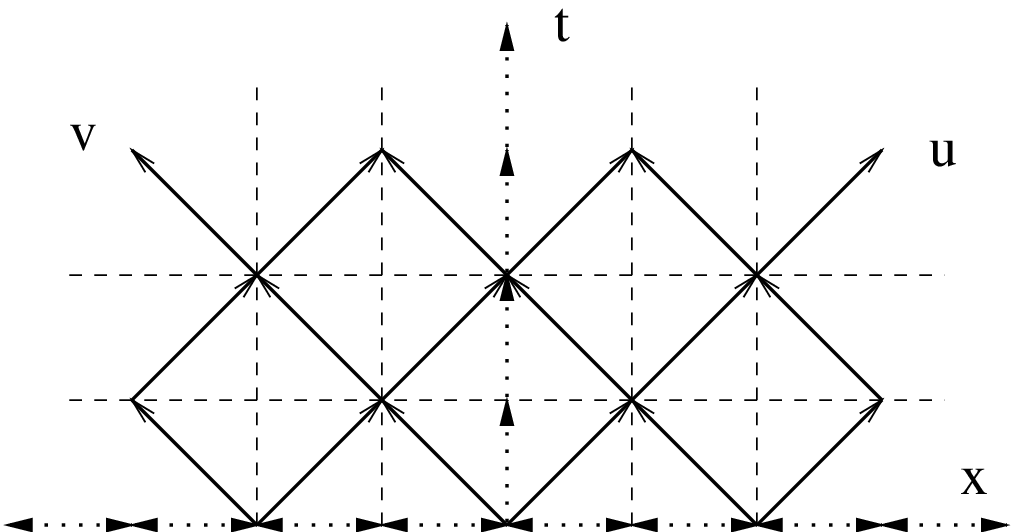}
\medskip

Coordinate transformation on the square lattice
\end{figure}

\vskip.2cm

\noindent This is the adjoint of Smoluchowski's equation for a
constant force field, of intensity proportional to $\gamma$ and
diffusion constant $h$, derived here in a conceptual framework
totally different from that conventionally used (e.g.\ the theory
of Markov processes; see \cite{Kac1956}, \S3).
\medskip

\noindent {\bf 2.} Evolution in velocity space.\\ Here $x$
represents the velocity of a particle and we take (cf.\
(\ref{4.3}))
\be
X^{\vec{k},\vec{k}+\hat{1}}= {1 \over 2} - \beta b (k^1-k^2),\quad
X^{\vec{k},\vec{k}+\hat{2}}= {1 \over 2} + \beta b (k^1-k^2) \, ,
\qquad
 \beta = \mathnormal{constant} >0
\label{4.8} \ee hence by (\ref{3.24}), (\ref{2.4})
\be
p= {1 \over 2} - \beta b (u-v) \, , \qquad q= {1 \over 2} + \beta
b (u-v) \, , \qquad p-q = -2 {b \over a} \beta x \label{4.9} \ee
so that in the continuous limit, (\ref{4.2}) gives
\be
\pa_t f+2\beta  x \pa_x f - {h \over 2} \pa_x^2 f  =0 \label{4.10}
\ee This is the adjoint to the 1D FP equation in velocity space,
giving the Ornstein-Uhlenbeck process (diffusion constant $h$,
drift coefficient $\beta$). We may remark however, that if $x$ is
interpreted as a position coordinate, then (\ref{4.10}) is the
adjoint to the Smoluchowski equation for a harmonic external field
(cf.\ \cite{Kac1956}, \S4).

\section{Evolution Equation: The General Case}
\setcounter{equation}{0}

In this section we apply the results of \S3 in the case of an
$(N+1)$-oriented hypercubic lattice $\M$.

The lattice coordinates $u^\mu$ are dimensionless. We consider
linear coordinate transformations to physical coordinates
\be
x^\mu=a_\mu \sum_\nu A^\mu_\nu u^\nu, \qquad\quad
\mu,\nu=0,1,\ldots,N \label{5.1} \ee in which $a_\mu$ are scaling
parameters with the dimensions of $x^\mu$. In view of \S3.4 we
require that
\be
x^0=t,\qquad\quad \mbox{hence} \quad a_0=-b,\quad A^0_\mu=1
\label{5.2} \ee and we shall eventually consider the continuous
limit
\be
a_\mu\to 0,\qquad {a_ia_j\over b}\to h_{ij},\qquad {a_i\over
b}\to\pm\infty,\qquad A^\mu_\nu\to\hat{A}^\mu_\nu \label{5.3} \ee
with the $A^\mu_\nu$ depending on the $a_\mu$'s, in general.
Eq\eq{5.3} is motivated by the limiting procedure considered in
\cite{DT1996,DT1997} and \S2 here. Nevertheless, its nature is
further explored in \S8. Here we only notice that, in contrast to
$(t,x^i)$, the $u^\mu$ are not defined in this limit. Moreover, to
avoid a too technical development, we restrict the discussion to
linear transformations with constant coefficients, thus making the
physical and conceptual aspects of our approach more clear. In
what follows, latin indices run from 1 to $N$, greek ones from 0
to $N$ and symbols with a hat denote quantities in the limit
(\ref{5.3}).

By \eq{5.1} we can write
\be
u^\mu =\sum_\nu B^\mu_\nu {x^\nu\over a_\nu},\qquad\qquad \sum_\nu
A^\mu_\nu B^\nu_\rho=\delta^\mu_\rho,\quad
B^\mu_\nu\to\hat{B}^\mu_\nu \label{5.4} \ee so that \eq{3.22},
\eq{3.25} become with the aid of \eq{2.9}
\be
\d f=\left(\pa_{-t}f(t,x)-{1\over2}\Delta f(t-b,x)\right)\d t+
\sum_i\bar{\pa}_if(t-b,x)\,\d x^i \label{5.5} \ee where
\be
\bar{\pa}_i f(t,x)&:=&{1\over a_i}\sum_\mu f(t,x^j+a_j A^j_\mu)
B^\mu_i, \label{5.5'}\\ \Delta f(t,x)&:=&{2\over b}\left(\sum_\mu
f(t,x^j+a_jA^j_\mu)B^\mu_0-f(t,x)\right) \label{5.5''} \ee and
\be
\begin{array}{c}
\d t\bu\d t=-b\,\d t,\qquad\qquad \d t\bu\d x^i=-b\,\d x^i\\[5mm]
\displaystyle{ \d x^i\bu\d x^j=-{a_ia_j\over b}\sum_\mu A^i_\mu
A^j_\mu B^\mu_0\,\d t +\sum_{\mu,m}{a_ia_j\over a_m} A^i_\mu
A^j_\mu B^\mu_m\,\d x^m}
\end{array}\label{5.6}
\ee With the interpretation of vector fields\  $X$ given in
\S\S3.2, 3.3, we consider those $X$ that satisfy \eq{3.26}, i.e.\
\be
X=\sum_\mu P^\mu \pa_{+u^\mu},\qquad\quad P^\mu\geq0,\quad
\sum_\mu P^\mu=1 \label{5.7} \ee hence $\langle \d t,
X\rangle=-b$, $\langle \d u^i,X\rangle=P^i$. Therefore the
evolution equation \eq{3.33} becomes (c.f. \ (\ref{4.2}))
\be
-{1\over b}X(f)=\pa_{-t}f(t,x)-{1\over2}\Delta f(t-b,x)-\sum_i
{a_i\over b}(AP)^i\,\bar{\pa}_if(t-b,x)=0 \label{5.8} \ee where,
by \eq{5.1}
\be
\langle\d x^i,X\rangle =a_i \,(AP)^i \label{5.11} \ee With the
interpretation of vector fields introduced in \S3, this is {\em
the average value of the ``infinitesimal'' change of $x^i$ along
$X$ in one time-step $b$}\/. By assuming that $f$ is sufficiently
well-behaved function, Taylor expansion of \eq{5.8} around $x^i$,
gives
\be
-{1\over b}X(f)=\pa_tf-\sum_i{a_i\over b}\,(AP)^i\,\pa_i
f-{1\over2} \sum_{i,j}{a_ia_j\over b}\left(\sum_\mu A^i_\mu
A^j_\mu P^\mu\right)\pa_i\pa_j f+o(a)=0 \label{5.9} \ee

We proceed further in analogy with conventional dynamics. Let
$(R^i)$ be the generator of motion in this case (e.g. in Newton's
or Hamilton's equations). Then, as it is explained in appendix
A.2, to the system of characteristic equations of the observables'
evolution equation (\ref{A3}iii), are associated the 1-forms
\be
\alpha^i = \d x^i + R^i \d t   \label{5.*} \ee which {\em vanish
along the integral curves of the vector field} that gives the
dynamical evolution.

Therefore, in the present discrete context, we {\em postulate} the
existence of dynamics in the above sense, that is, the generator
of dynamical evolution is determined by a similar relation, namely
\be
\langle \alpha^i , X \rangle = \langle\d x^i,X\rangle + R^i
\langle \d t , X \rangle =0 \quad \mbox{hence} \quad a_i(AP)^i= b
R^i \label{5.**} \ee (cf.\ (\ref{A4'}). Hence,
\be
(AP)^i= {b \over a_i} R^i = {a_i\over h_{ii}} R^i + o(a) \qquad
\mbox{with}\qquad R^i\to\hat{R}^i \label{5.12} \ee Notice that the
postulated existence of dynamics, that is (\ref{5.**}i), implies
that $\langle \d x^i, X \rangle$ is $o(b)$, thus justifying this
assumption which was introduced in the derivation of (\ref{4.4}).
It is also equivalent to a similar relation determining the
transition probabilities $P^\mu$, as explained in appendix A.3.

Since $(AP)^0=1$ (because of \eq{5.2}, \eq{5.7}), we can invert
\eq{5.12} to get
\be
P^\mu=B^\mu_0+\sum_m{a_m\over h_{mm}}\,B^\mu_m
R^m,\qquad\mbox{hence in the limit}\qquad\hat{P}^\mu=\hat{B}^\mu_0
\label{5.13} \ee so that by \eq{5.7}
\be
\sum_\mu \hat{B}^\mu_0=1,\qquad\qquad \hat{B}^\mu_0\geq0
\label{5.14} \ee Therefore \eq{5.9} can be rewritten as
\be
-{1\over b}X(f)=\pa_tf-\sum_i{a_i{}^2\over h_{ii}b}\,R^i\,\pa_i
f-{1\over2} \sum_{i,j}\eta^{ij}\pa_i\pa_j f+o(a)=0 \label{5.15}
\ee where
\be
\eta^{ij}:={a_ia_j\over b}\sum_\mu A^i_\mu A^j_\mu B^\mu_0\quad
\to\quad \hat{\eta}^{ij}=h_{ij}\sum_\mu \hat{A}^i_\mu
\hat{A}^j_\mu\hat{P}^\mu \label{5.16} \ee In the continuous limit
\eq{5.3}, we have
\be
\pa_tf-\sum_i\hat{R}^i\pa_if-{1\over2}\sum_{i,j}\hat{\eta}^{ij}\pa_i\pa_j
f=:\hat{X}f=0 \label{5.17} \ee This is our general evolution
equation for observables $f$, corresponding to the DC obtained
from \eq{5.6} in the limit \eq{5.3}
\be
\d t\bu\d t=0,\qquad \d t\bu\d x^i=0,\qquad\d x^i \bu \d
x^j=-\hat{\eta}^{ij}\d t \label{5.18} \ee Eq\eq{5.18} is identical
with the DC considered in \cite{DT1996,DT1997}, called 2nd order
DC there, because it leads to 2nd order evolution equations, i.e.\
\eq{5.17}. We shall come to this point again in \S9.

A number of interesting remarks can be made here:
\smallskip

\noindent 1) If \eq{5.14} holds, then $\hat{\eta}^{ij}$ is a
non-negative definite matrix, so that \eq{5.17} is the adjoint of
a generalized FP-equation with diffusion matrix
$\hat{\eta}^{ij}$and drift vector $\hat{R}^i$.
\smallskip

\noindent 2) Under the transformation \eq{5.1}, the correlation
matrix of the lattice, \eq{3.27}, becomes
\be
H^{\mu\nu}:=(A\mathbb{P}A^t)^{\mu\nu} \label{5.19} \ee
\be
H^{\mu0}&=&0 \nonumber\\ H^{ij}&=&\sum_\mu a_ia_j A^i_\mu A^j_\mu
P^\mu-a_ia_j(AP)^i(AP)^j= \langle\d x^i\bu\d
x^j,X\rangle-\langle\d x^i,X\rangle\langle\d x^j,X\rangle
\nonumber\\ &=&\langle\d(x^ix^j)-x^i\d x^j-x^j\d
x^i,X\rangle-\langle\d x^i,X\rangle\langle\d x^j,X\rangle
\label{5.20} \ee where we have used the Leibniz rule \eq{1.5}.
\begin{itemize}
\item $H^{\mu0}=0$ reflects the fact that the time change $\d t$ is the
(essentially) unique eigenvector of zero of $\mathbb{P}$, as
discussed in \S3.4
\item Because of \eq{5.12}, the last term on the r.h.s of \eq{5.20} is
$o(a_i{}^4)$.
\item Using the unit $\rho$ of the algebra $(\O^1,\bu)$,
eq(\ref{3.29}), we may write \bez H^{ij}= \langle \left( \d x^i -
\langle \d x^i,X \rangle \rho \right) \bu \langle \left( \d x^j -
\langle \d x^j,X \rangle \rho \right),X \rangle \eez
(cf.(\ref{3.31'})). In fact, $H^{ij}$ is the correlation matrix of
the 1-forms $\alpha^i$, eq(\ref{5.*}), associated with the
equations of motion via (\ref{5.**})
\be
H^{ij}&=& \langle \left( \alpha^i - \langle \alpha^i,X \rangle
\rho \right) \bu \langle \left( \alpha^j - \langle \alpha^j,X
\rangle \rho \right),X\rangle \nonumber \\ &=&\langle \alpha^i \bu
\alpha^j,X \rangle - \langle \alpha^i,X \rangle \langle \alpha^j,X
\rangle \nonumber  \\ &=& \langle \alpha^i \bu \alpha^j,X \rangle
\label{5.22'} \ee
\end{itemize}
Therefore in the continuous limit we get
\be
{H^{ij}\over b}\;\;\to\;\; \hat{\eta}^{ij}&=&\langle\d x^i\bu\d
x^j,\hat{X}\rangle =\langle\d(x^ix^j)-x^i\d x^j- x^j\d
x^i,\hat{X}\rangle \nonumber\\ &=&
\hat{X}(x^ix^j)-x^i\hat{X}(x^j)-x^j\hat{X}(x^i) \label{5.21} \ee
Thus, $\hat{\eta}^{ij}$ is essentially the limiting value of (the
matrix of) correlations of ``infinitesimal'' changes along the
axes of the lattice, induced by the generator of evolution $X$. In
fact, by \eq{5.19}, $\hat{\eta}^{ij}$ vanishes if $X$ is the
generator of a deterministic flow on the lattice, i.e.\ when the
concept of a trajectory is well-defined for $X$, as explained in
\S3.3 (cf.\eq{3.18}). This is reflected in the limiting equation
\eq{5.21}, which gives the deviation of $\hat{X}$ from an ordinary
derivation, that is, an ordinary vector field\ for which the
concept of a trajectory is (locally) always meaningful. We may
also compare $\hat{\eta}^{ij}$ with the usual interpretation of
the diffusion matrix in the kinetic theory of open systems. There,
it is the correlation matrix of the hamiltonian vector field\ of
the interaction hamiltonian of the open system with the bath, with
respect to the state of the bath (see e.g.\ \cite{GrT1988}
eq(4.16)). A similar interpretation of this quantity exists in the
theory of markovian stochastic processes and of It\^o's stochastic
calculus (e.g.\ \cite{Gard1985} chs.3, 4, \cite{vKamp1981},
ch.VIII).

On the other hand, the above comments provide an interpretation of
the noncommutativity of the DC defined by \eq{5.18} and studied in
previous works (\cite{DT1996,DT1997}): Starting with a {\em
discrete structure} of the phase-space, somehow reflects the fact
that originally one considers the {\em fine} microscopic structure
of the system. At this level, changes are described by structures
that are {\em necessarily noncommutative} and it is a {\em
generic} characteristic of vector fields\ in this case, {\em not}
to describe deterministic motion (i.e.\ flows of trajectories), in
contrast with vector fields\ in the (ordinary) continuous case for
which this is {\em always} the case. Then, the passage to the
continuous limit corresponds to a coarse-graining process, in
which the fine details of the micro-structure are no longer
explicit, but which, nevertheless, have been taken into account.
In particular, the above generic characteristic of vector fields\
on a discrete phase-space is expressed quantitatively in the
continuous limit, by the non-vanishing of $\hat{\eta}^{ij}$,
eq\eq{5.16}.
\smallskip

\noindent 3) In 1) above, it was mentioned that
$(\hat{\eta}^{ij})$ is nonnegative-definite (being the limit of
such matrices). On the other hand it is an easy matter to see that
for such a matrix, $\hat{\eta}^{ii}=0$ for some fixed $i$ implies
$\hat{\eta}^{ij}=0$ for all $j$.\footnote{$(\hat{\eta}^{ij})$
defines a (possibly degenerate) nonnegative-definite scalar
product on $\mathbb{R}^N$, and we apply the Schwartz inequality to
the $i$-th element of the canonical basis of $\mathbb{R}^N$ with
each $j$-th element $j\neq i$.} In our case, this means that with
respect to $x^i$, $\hat{X}$ in \eq{5.17} is a derivation, hence it
describes deterministic motion along the $x^i$-axis.
\smallskip

\noindent 4) The limiting eq\eq{5.17} is the most general linear
autonomous differential (evolution) equation, whose adjoint admits
a probabilistic interpretation; more precisely, under mild
regularity conditions on the coefficients, it defines a positivity
and normalization preserving, strongly continuous semigroup,
globally on the space of (continuous) observables $f$ having a
finite limit at infinity. Its adjoint also defines a positivity
and normalization preserving semigroup, satisfying a general
$H$-theorem (\cite{TGr1999}, theorems 5.1, 6.1). This is assured
by the non-negative definiteness of $\hat{\eta}^{ij}$. In our case
this follows from (\ref{5.14}), which is a consequence of the
interpretation of a vector field\ as a transition
probability-distribution for infinitesimal changes. Thus, if
$\{\tilde{\pa}_t,\tilde{\pa}_{x^i}\}$ is the basis of vector
fields\ dual to $\{\d t,\d x^i\}$, then \bez B^\mu_0=\langle\d
u^\mu,-b\tilde{\pa}_t\rangle=-b\tilde{\pa}_t(u^\mu) \quad \to
\quad \hat{B}^\mu_0\geq 0 \eez Therefore, $\hat{B}^\mu_0\geq0$
means that in the {\em limit of the continuum} (i.e.\ in a
macroscopic description --- see 2) above), changes in phase-space
are in the {\em future} direction (cf.\ the discussion following
eq\eq{A3} in Appendix A).  Thus, in this picture, the ``
macroscopic'' direction of time is intimately related to a
probabilistic interpretation of evolution at the ``microscopic''
level (here the terms ``macroscopic'' and ``microscopic''
correspond to the terms ``continuous description'' and ``discrete
description'' in the sense of 2) above).
\smallskip

\noindent 5) The expansion of the evolution equation \eq{3.33} (or
\eq{5.8}) in powers of the $a_i$ is not essential, but has been
done because of the generality of the transformation \eq{5.1}. In
each particular case $X(f)$ may be written explicitly in terms of
appropriate difference operators, which in the continuous limit
yield the corresponding FP equation (e.g.\ as in \S4). Such an
$N$-dimensional example is worked out in Appendix B.

\section{Transformations in phase-space}
\setcounter{equation}{0}

There is considerable freedom in the choice of the coordinates
$(x^i)$, eq\eq{5.1}, in  such a way that the continuous limit
exists. In this section we comment briefly on this issue.

Let $(x^\mu)$, $(x'{}^\mu)$ be coordinate systems on $\M$,
obtained from the lattice coordinates by a linear transformation
like eq\eq{5.1} and let $x'{}^\mu=\Lambda^\mu_\nu x^\nu$. In view
of the discussion in \S\S3.3, 3.4, 5, from a physical point of
view we require that $\Lambda$ leaves $\d t$ invariant. Then it
has the form
\be
\Lambda=\left(\begin{array}{c|c} 1 & \mathbf{0}\\ \hline
\\[-4mm]\Lambda^i & \Lambda^i_j\end{array}\right) \ee Such
transformations form  a group. Moreover, since $\d t$ is in the
kernel of the correlation matrix $(H^{\mu\nu})$, eq\eq{5.19},
$\Lambda$ leaves the form of $H^{\mu\nu}$ (i.e.\ $H^{\mu 0}=0$)
unaltered:
\be
H=\left(\begin{array}{c|c} 0 & \mathbf{0}\\ \hline \\[-4mm] 0 &
H^{ij}\end{array}\right)\quad
\stackrel{\Lambda}{\longrightarrow}\quad
H'=\left(\begin{array}{c|c} 0 & \mathbf{0}\\ \hline \\[-4mm] 0 &
(\Lambda H\Lambda^t)^{ij}\end{array}\right) \ee If we write
\eq{5.1} in the form
\be
(u^\mu)\;\stackrel{A}{\longrightarrow}\;\left({x^\mu\over
a_\mu}\right):\qquad \qquad A=\left(\begin{array}{c|c} 1 &
\mathbf{1}\\ \hline \\[-4mm] A^i & A^i_j\end{array}\right) \ee
where $\mathbf{1}=(1,1,\ldots,1)$, then
\be
H'=(\Lambda A)\mathbb{P}(\Lambda A)^t \ee A natural (partial)
fixing of the coordinate freedom is to choose $\Lambda$ so that
the correlation matrix, hence $\hat{\eta}^{ij}$ in \eq{5.17}, is
diagonal. However  many other possibilities exist, see e.g.\
Appendix B and section 7.

\section{Further Applications: 2D Evolution Equations}
\setcounter{equation}{0}

As a concrete application of the results of \S5, which also
exhibits the main features of the general $N$-dimensional case, we
consider in this section the general form of the evolution
equation (\ref{5.17}) in two dimensions. Then, we show how
evolution equations that are used in statistical physics as model
kinetic equations can be derived in the present context, at the
same time throwing some light into their foundations from a new
perspective.

An example of this type is Kramers' equation which for simplicity
is considered here in one spatial dimension $x$:
\be
\pa_t \rho + y \pa_x \rho +F(x) \pa_y \rho - \beta \pa_y(y \rho)
-{h\over 2} \pa_y^2 \rho =0 \label{7.1} \ee This is a kinetic
equation for the probability distribution $\rho$ of a (unit mass)
brownian particle of velocity $y$ in an equilibrium bath of
temperature $h/2\beta$, $\beta$ being its friction (drift)
coefficient and $F$ an external force field.

Eq(\ref{7.1}) can be derived from microscopic dynamics, e.g.\ by
expanding the so-called generalized master equation in powers of
the ratio of the masses of the bath particles to that of the
brownian particle, for brevity taken here equal to 1 (see e.g.\
\cite{RedeL1978}, ch.IX.4, \cite{Mazo1978} \S5). We notice
however, that (\ref{7.1}) has been extrapolated far beyond the
usual model of brownian motion, namely, a heavy particle
interacting via {\em hard collisions} with a bath of much lighter
particles (e.g.\ it has been used for self-gravitating systems,
see the discussion in \cite{TGr1988} \S5 and references therein).

On the other hand, (\ref{7.1}) is also obtained by using
Langevin's equation as a dynamical model and assuming, that the
motion of the particle is a Markov process, and that {\em motion
along $x$ is deterministic} so that the dependence of the
transition probability on $x$ is a $\delta$-function (see e.g.\
\cite{C1943} eq(242), \cite{RedeL1978} ch.II.2, \cite{vKamp1981}
\S VIII.7). This is an {\em extra condition}, the use of which is
reflected in the derivation of (\ref{7.1}) in the context of StC
by writing Langevin's equation as a stochastic differential
equation
\be
\d x = y \d t \,  \quad\qquad \d y = -( \beta y - F(x)) \d t + h
\d W_t \label{7.2} \ee where $W_t$ is a standard Wiener process
and {\em no stochastic term appears in} $\d x$ (see e.g.\
\cite{Arn1973}, theorem 9.3.1, \cite{Gard1985}, \S5.3.6,
\cite{Blanch1987} \S II.3). This extra condition is often
justified by saying that the external field varies slowly on a
length scale in which the velocity is damped (e.g.\
\cite{vKamp1981}, p.232). Nevertheless, there is no generally
accepted view on how (\ref{7.1}) should be generalized if this
extra condition is relaxed, although second order derivatives in
$x$ are expected to appear on the basis of a systematic analysis
of the microscopic dynamics of realistic classical 3D-models of
brownian particles weakly coupled to the bath (\cite{T1988} \S2,
\cite{FriGo1984} \S4).\footnote{The same conclusion follows from a
similar analysis of quantum systems and by taking the classical
limit of the resulting equations; e.g.\ for a harmonic oscillator
in a bath of oscillators, such an analysis leads to a widely used
kinetic equation of the Lindblad type (i.e.\ conserving density
matrices), whose classical limit is a FP equation involving a
drift and a diffusion term in the $x$ axis (\cite{TGrHa1998} \S\
3,4).} Such 2nd order derivatives are not easily accounted for by
approaches based on StC, since this presupposes the appearance of
a stochastic term in the first of eqs(\ref{7.2}), not easily
justified from a physical point of view.

On the other hand, in \cite{DT1996} \S5 we have shown how such
evolution equations (including (\ref{7.1}) are formally
incorporated as hamiltonian equations in symplectic mechanics
developed in the context of the noncommutative DC defined
by(\ref{5.18}). Below we explore the physical interpretation of
this formal result by employing the techniques and conceptual
framework of the previous sections.

In the notation of \S5 we write for a (2+1)D-system
\be
(u^\mu) = (u,v,w) \, , \;\; (x^\mu) = (t,x,y) \, , \;\; (a_i) =
(a, \bar{a}) \, , \; \; P^\mu = (p,q,r) \, , \;\; X=\sum_\mu P^\mu
\pa_{+u^\mu} \label{7.3} \ee
\be
(A^\mu_\nu) = \left( \matrix{ 1 & 1 & 1\cr \kappa & \lambda & \mu
\cr \kappa' & \lambda' & \mu'} \right) \label{7.4} \ee Then, by
(\ref{5.11}), (\ref{5.12}), (\ref{5.16}), we have
\be
X(x) = \langle \d x , X \rangle &=& a(\kappa p + \lambda q + \mu
r) = {a^2 \over h_{11}} R_x  \label{7.5}\\ X(y) = \langle \d y , X
\rangle &=& \bar{a}(\kappa' p + \lambda' q + \mu ' r) = {\bar{a}^2
\over h_{22}} R_y  \label{7.6} \ee
\be
\eta^{11} = {a^2 \over b} (\kappa^2 p + \lambda^2 q + \mu^2
r),\qquad \eta^{12} = {a \bar{a} \over b} (\kappa \kappa' p +
\lambda \lambda' q + \mu \mu'r) \label{7.7} \ee
\be
\eta^{22} = {\bar{a}^2 \over b} (\kappa'{}^2 p + \lambda'{}^2 q +
\mu'{}^2 r) \label{7.8} \ee where the limiting probabilities in
the continuous limit (\ref{5.3}) are readily found via
(\ref{7.4}), (\ref{5.13}) to be
\be
(\hat{p},\hat{q},\hat{r}) = \left({\hat{\lambda} \hat{\mu'}-
\hat{\lambda'}\hat{\mu} \over |\hat{A}|} \, , {\hat{\kappa'}
\hat{\mu}- \hat{\kappa}\hat{\mu'} \over |\hat{A}|} \, ,
{\hat{\lambda '} \hat{\kappa}- \hat{\lambda}\hat{\kappa'} \over
|\hat{A}|} \right) \; , \qquad  |\hat{A}|= \det(\hat{A}_\nu^\mu )
\label{7.9} \ee Finally, the evolution equation (\ref{5.15}) is
\be
\pa_tf - {a^2 \over b h_{11}} R_x \pa_xf - {\bar{a}^2 \over b
h_{22}} R_y \pa_xf -{1 \over 2} \left( \eta_{11} \pa_x^2 f +
\eta_{12} \pa_{xy}^2 f +\eta_{22}\pa_y^2 f\right) + o(a,\bar{a}) =
0 \label{7.10} \ee Now suppose that $(x,y)$ are the phase space
coordinates of a particle, moving under friction linear in the
velocity $y$, say $-\beta y$, and an external field $F(x)$.
Evidently, the generator of {\em newtonian motion} is $(y, -(\beta
y -F(x))$ and therefore as in \S5 (cf. (\ref{5.*})) we may
introduce the 1-forms \bez \alpha_1 = \d x +y \d t  \, , \qquad
\alpha_2 = \d y - (\beta y - F(x)) \d t \eez Consequently
(\ref{5.**}), giving the ``infinitesimal'' changes  of $(\d x, \d
y)$ in time $\d t$, becomes $\langle \alpha_i \, , \, X \rangle =
0, \; i=1,2$, hence
\be
\langle \d x \, , \, X \rangle = b y = b R_x \, , \qquad \langle
\d y \, , \, X \rangle = -b( \beta y -F(x)) = b R_y \label{7.11}
\ee and in the limit
\be
R_x = \hat{R}_x =y \, , \qquad\qquad R_y = \hat{R}_y = -(\beta y -
F(x)) \label{7.12} \ee If {\em motion along $x$ is deterministic
in the continuous limit}, that is, the concept of a trajectory in
$x$ is meaningful, then by the remark (3) in \S5, this is
equivalent to
\be
\hat{\eta}_{11}=0 \;\;\; (\mbox{hence} \; \hat{\eta}_{12} =0),
\quad\quad \mbox{that is} \quad\quad \hat{\kappa}^2 \hat{p}+
\hat{\lambda}^2 \hat{q}+\hat{\mu}^2 \hat{r}=0
 \label{7.13}
\ee This corresponds to the absence of a stochastic term in
(\ref{7.2}i). Since $\hat{p}\, , \hat{q}\, , \hat{r}$ are
nonnegative, it is readily obtained from (\ref{7.13}), (\ref{7.9})
that up to a permutation of the lattice coordinates $(u,v,w)$
(hence of $(p,q,r)$), there are two possibilities:
\medskip

\noindent {\bf (1)}  $\hat{q} = \hat{r}=0$, $\hat{p}=1$,
$\hat{\kappa}=\hat{\kappa'} =0$ implying  $\hat{\eta}_{22}=0$,
 $\hat{\lambda}\hat{\mu
'} - \hat{\lambda '}\hat{\mu}
 \neq 0$.
Then, (\ref{5.17}), the evolution equation in the continuous
limit, is
\be
\pa_t f - y \pa_x f +(\beta y - F(x)) \pa_y f=0 \label{7.14} \ee
This describes motion with velocity damping and well defined
trajectories in {\em phase space}\/. There is a 4-parameter gauge
freedom for the transformation (\ref{7.4}).
\medskip

\noindent {\bf (2)}
\be
\quad \hat{q} = 0 \, , \;\; \hat{p} ={\hat{\mu}' \over
\hat{\mu}'-\hat{\kappa}'}
 \, , \;\; \hat{r} ={\hat{\kappa}' \over
\hat{\kappa}'-\hat{\mu}'}\, , \quad
 \kappa=\mu=0 \, , \quad \hat{\eta}_{22} =
-\hat{\kappa}'\hat{\mu}'>0
 \label{7.15}
\ee with the evolution equation (\ref{5.17}) being
\be
\pa_t f - y \pa_x f +(\beta y - F(x)) \pa_y f +{1 \over 2}
(h_{22}\hat{\kappa}'\hat{\mu}')\, \pa_y^2 f =0 \label{7.16} \ee
and there is again a 4-parameter gauge freedom subject to the
constraint $\hat{\kappa}'\hat{\mu}'<0$. This is formally identical
with the adjoint of (\ref{7.1}). This formal similarity is more
than a coincidence and gives another illustration of the
consistency of the probabilistic framework introduced in \S3 and
elaborated in \S5. A simple choice of the gauge is
$\kappa'=-\mu'=1$, $\lambda'=0$. Then, in view of (\ref{7.4}),
(\ref{5.13}), eqs(\ref{5.16}), (\ref{5.20}) imply \bez \eta^{22}=
{\bar{a}^2 \over b} (p+r) + o(a, \bar{a}), \qquad {1 \over b}
\langle \d y \bu \d y , X \rangle = \eta^{22} + o(a, \bar{a}) \eez
Therefore \bez {1 \over b} \langle \d y \bu \d y , X \rangle =
{\bar{a}^2 \over b} (p+r) +o(a, \bar{a}) = {\bar{a}^2 \over b}
(1-q) +o(a, \bar{a}) = {\bar{a}^2 \over b}+o(a,\bar{a},q) \to
\hat{\eta}^{22}=h_{22} \eez i.e.\ $\bar{a}^2/b$ is the average of
the square of the velocity change from $y$ to $y+ \d y$ in a
``small'' time interval $b$ divided by $b$.

On the other hand, it is well known that in the usual
interpretation of (\ref{7.1}) as a model of brownian motion, $h =
\beta \bar{y}^2$, with $\bar{y}=$thermal velocity, whereas, for
{\em short} times $t$ in which the velocity changes from $y_0$ to
$y$, $\langle (y-y_0)^2 \rangle /t = \beta \bar{y}^2 +o(t) = h
+o(t)$ with $\langle , \rangle$ denoting the average over the bath
(see e.g. \cite{C1943} eq(161)). Thus, in the continuous limit
$h_{22} \equiv h$ and (\ref{7.16}) coincides with the adjoint of
(\ref{7.1}).

It should be noticed here, that this result follows by {\em
assuming} that in the configuration space (i.e.\ in the $x$-axis),
motion is along well-defined trajectories. It is a remarkable fact
however, that if this assumption is relaxed, then (\ref{7.10})
reduces to an equation describing diffusion in the $x$-axis as
well, a fact not easily accommodated in other approaches, as
explained at the beginning of this section.
\medskip

\noindent {\bf Remarks}: (a) In the present section and in \S4, we
have seen that up to a coordinate transformation, our approach
allows for concrete results in specific cases, once the drift term
$\langle \d x^i,X \rangle$ i.e.\ the average ``infinitesimal''
change of $x^i$ is given, that is, once a prescription is given
for the choice of $X$ in each particular case. In \cite{DT1996},
$X$ was specified by assuming it to be hamiltonian in the context
of the ``second order calculus'' (\ref{5.18}). This points out to
the need of developing symplectic mechanics on the oriented
hypercubic lattice, to be presented in another paper.

(b) A similar, but simpler procedure yields in the 1D case the
following: We write (\ref{5.1}), (\ref{5.3}) as \bez t=-b(u+v) \,
, \qquad x= a (\kappa u + \lambda v) \, , \qquad {a^2 \over b}
\longrightarrow h \eez Then using (\ref{4.1}), eq(\ref{5.13})
becomes \bez \hat{p}= - {\hat{\lambda} \over \hat{\kappa} -
\hat{\lambda}} \, , \qquad \hat{p}= {\hat{\kappa} \over
\hat{\kappa} - \hat{\lambda}} \, , \qquad \hat{\lambda}
\hat{\kappa}<0 \eez and the evolution equation in the continuous
limit, eq(\ref{5.17}), gives \bez \pa_t f- \hat{R} \pa_x f + {1
\over 2} h (\hat{\lambda} \hat{\kappa}) \pa_x^2 f = 0 \eez Notice
that $\langle \d x , X \rangle /b = a (\kappa p + \lambda q)/b =
a^2 R/b h \longrightarrow \hat{R}$ so that the results of \S4 are
recovered if $\hat{\kappa}=- \hat{\lambda} =1$ (cf.\
eqs(\ref{4.6})-(\ref{4.10})).

\section{Comments on higher order equations}
\setcounter{equation}{0}

The basic methodological ``rules'' of the present approach
introduced in the previous sections are:\begin{itemize}
\item The extended phase-space has the structure of a (oriented hypercubic)
lattice $\M$.
\item The phase space coordinates $(x^\mu)$ are obtained by a coordinate
transformation from the lattice coordinates $(u^\mu)$, involving
the scaling parameters $(a_\mu)$, so that the usual continuum
description in terms of the $(x^\mu)$ is recovered in the limit
\eq{5.3} (cf. \cite{Roe1994}, p.244).
\end{itemize}
In principle however, one can imagine other limits as well. In
this section we first examine this possibility in a somewhat more
general setting and then we reconsider the case of the oriented
hypercubic lattice in the light of the results obtained.

Let the DC on $\M$ be defined by
\be
\d u^\mu\bu\d u^\nu = C^{\mu\nu}_\rho \d u^\rho, \qquad\qquad
C^{\mu\nu}_\rho\in \A \label{8.1} \ee where  the summation
convention has been used (it is used throughout this section).
This may be considered as a class of (algebraic) deformations of
the ordinary DC. It can be shown that (\cite{BDM-H1995} \S4.5)
\bez \d f= D_\rho f\d u^\rho, \qquad\qquad f=f(u)\in\A \eez
\be
D_\rho f =\pa_\rho f+\sum_{r=2}^{+\infty} {1\over
r!}\sum_{\mu_1,\ldots,\mu_r \atop \nu_1,\ldots,\nu_{r-2}}
C^{\mu_1\mu_2}_{\nu_1}C^{\mu_3\nu_1}_{\nu_2}\cdots
C^{\mu_r\nu_{r-2}}_\rho\,\pa_{\mu_1}\cdots\pa_{\mu_r} f
\label{8.2} \ee We pass to the continuous limit by dividing
$\{u^\mu\}$ into several groups, with the same scaling in each
group. In each case one of the groups contains the ``time''
variable(s). It will be readily seen, that the results obtained
are valid even if in each group, each coordinate is scaled
differently, provided the order of magnitude is the same for all
elements of the same group. Moreover, any finite limit of the
scaling parameters is set equal to 1, to avoid useless cumbersome
notation.
\bigskip

\noindent {\bf (1) Division of $\{u^\mu\}$ into two groups:} \bez
u^a=:{t^a\over b},\quad a=0,\ldots,n \:,\qquad u^i=:{x^i\over a},
\quad i=n+1,\ldots,N \eez continuous limit:
\be
a,b\to 0\,, \qquad {a^2\over b}\to 1\label{8.3} \ee From \eq{8.1}
and \eq{8.2} we get
\be
\begin{array}{rcl}
\d t^a\bu\d t^b &=& b \,C^{ab}_c\,\d t^c+\displaystyle{b^2\over a}
\,C^{ab}_i\,\d x^i,\\[3mm] \d t^a\bu\d x^i &=& a\, C^{ai}_b\,\d
t^b+ b\, C^{ai}_j\, \d x^j,\\ \d x^i\bu\d x^j &=&
\displaystyle{a^2\over b}\,C^{ij}_a\,\d t^a+a\,C^{ij}_k\,\d x^k
\end{array} \label{8.4} \ee
\be
\begin{array}{rcl}
D_a f &=& \pa_a f+\displaystyle{1\over2}b \,C^{bc}_a\,\pa_b\pa_c f
+a\,
  C^{bi}_a\,\pa_b\pa_i f +\displaystyle{{1\over2}{a^2\over b}}\,
C^{ij}_a\,\pa_i\pa_j f +o(a)\\[3mm] D_i f &=& \pa_i f +
\displaystyle{{1\over2}{b^2\over a}} \,C^{ab}_i\,\pa_a\pa_b f+
b\,C^{aj}_i\,\pa_a\pa_j f+\displaystyle{1\over2}
a\,C^{jk}_i\pa_j\pa_k f + o(a)
\end{array}
\label{8.5} \ee If as in \S5, quantities with a hat denote values
in the limit \eq{8.3}, then in that limit, \eq{8.4}, \eq{8.5}
reduce to (cf. eqs\eq{5.6}, \eq{5.18} and \cite{DT1996} eq.(4.13))
\be
\d t^a\bu\d t^b=0,\qquad \d t^a\bu\d x^i=0,\qquad \d x^i\bu\d
x^j=\hat{C}^{ij}_a\d t^a \ee
\be
D_a f=\pa_a f+{1\over2}\hat{C}^{ij}_a\,\pa_i\pa_j f,\qquad\qquad
D_i f= \pa_i f \ee so that the evolution equation \eq{3.33}
reduces to a well-defined equation of the form \eq{5.17}.
\bigskip

\noindent {\bf (2) Division of $\{u^\mu\}$ into three groups:}
\bez u^a=:{t^a\over b},\;a=0,\ldots,m\,,\qquad u^r=:{y^r\over
c},\;r=m+1,\ldots,n\,,\qquad u^i=:{x^i\over a},\;i=n+1,\ldots,N
\eez continuous limit:
\be
a,b,c\to 0,\qquad {a^2\over c}\to 1,\qquad {a^3\over b}\to 1
\label{8.8} \ee Eqs \eq{8.1}, \eq{8.2} imply
\be
\begin{array}{rcl}
\d t^a\bu\d t^b &=& \displaystyle{b\,C^{ab}_c \d t^c +{b^2\over
c}\,C^{ab}_r \d y^r +{b^2\over a}\, C^{ab}_i\d x^i }\\ \d t^a\bu\d
y^r &=& \displaystyle{c\,C^{ar}_b\d t^b+b\,C^{ar}_s\d y^s+{bc\over
a}\,C^{ar}_i\d x^i }\\ \d t^a\bu\d x^i &=&
\displaystyle{a\,C^{ai}_b\d t^b + {ab\over c}\,C^{ai}_r \d
y^r+b\,C^{ai}_j\d x^j}\\ \d y^r\bu\d y^s &=&
\displaystyle{{c^2\over b}\,C^{rs}_a\d t^a+c\,C^{rs}_t\d
y^t+{c^2\over a}\,C^{rs}_i\d x^i }\\[3mm] \d y^r\bu\d x^i &=&
\displaystyle{{ac\over b}\,C^{ri}_a\d t^a+a\,C^{ri}_s\d y^s +
c\,C^{ri}_j\d x^j}\\[2mm] \d x^i\bu\d x^j
&=&\displaystyle{{a^2\over b}\,C^{ij}_a\d t^a+{a^2\over
c}\,C^{ij}_r\d y^r+a\,C^{ij}_k\d x^k}
\end{array}\label{8.9}
\ee
\be
D_i f=\pa_i f+o(a),\qquad\qquad D_r f=\pa_r f+{1\over2}{a^2\over
c}\,C^{ij}_r\pa_i\pa_j f+o(a^3) \label{8.10} \ee
\be
\lefteqn{D_a f = \pa_a f+{1\over2}{a^2\over
b}\,C^{ij}_a\pa_i\pa_jf + {ca\over b}\,
C^{ri}_a\pa_r\pa_if+}\hspace*{4cm}&& \nonumber\\ &&{1\over6}\left(
C^{il}_a C^{jk}_l+C^{ir}_a C^{jk}_r + C^{ib}_a
C^{jk}_b\right)\pa_i\pa_j\pa_kf +o(a) \label{8.11} \ee It is now
clear that in the limit \eq{8.8}, eqs\eq{8.10} and all but the
last one of \eq{8.9} have a well-defined limit. The coefficient of
$C^{ij}_a$ in
 the last of \eq{8.9} and in \eq{8.11} diverges as $1/a$.
Therefore, the limit
 \eq{8.8} exists in this case {\em only if we impose an extra
condition} on the
 structure constants which define the DC
\be
C^{ij}_a= a\, K^{ij}_a,\qquad\qquad K^{ij}_a\to\hat{K}^{ij}_a
\label{8.12} \ee In this case \eq{8.9}--\eq{8.11} reduce in the
continuous limit \eq{8.8}:
\be
\begin{array}{c}
\d t^a\bu\d t^b=0,\quad \d t^a\bu\d y^r=0,\quad \d t^a\bu\d
x^i=0,\quad \d y^r\bu\d y^s=0,\\[2mm] \d y^r\bu\d x^i
=\hat{C}^{ri}_a\d t^a,\qquad \d x^i\bu\d x^j=\hat{K}^{ij}_a\d
t^a+\hat{C}^{ij}_r\d y^r
\end{array}
\ee
\be
\begin{array}{c}\displaystyle{
D_if=\pa_if,\qquad D_rf=\pa_rf+{1\over2}\,\hat{C}^{ij}_r\pa_i\pa_j
f}\\[3mm] \displaystyle{ D_a f =
\pa_af+\hat{C}^{ri}_a\pa_r\pa_if+{1\over2}\hat{K}^{ij}_a
\pa_i\pa_jf+{1\over6}\hat{C}^{ir}_a\hat{C}^{jk}_r\pa_i\pa_j\pa_kf}
\label{8.14}
\end{array}
\ee so that the evolution equation \eq{3.33} reduces to an
equation containing 3rd order derivatives.

It is easily seen that proceeding in this way, that is, by
dividing $\{u^\mu\}$ into disjoint subsets, scaled by parameter
$a_A,\,A=1,2,\ldots,L$ and then considering the limits
\be
a_A\to 0,\qquad {(a_1)^A\over a_A}\to \mbox{finite},\qquad
C^{\mu\nu}_\rho\to\mbox{finite}  \label{8.15} \ee then unless
$L=2$, {\em extra conditions on $C^{\mu\nu}_\rho$ should be
imposed} so that in that limit the commutation relations and $\d
f$, $f\in\A$, are well-defined.

The significance of this general result can be better appreciated
if we implement the procedure described above, in the case of the
oriented hypercubic lattice $\M$, that is, when the DC is that of
\S5, \eq{5.6}. Instead of the limit \eq{5.3} (or equivalently
\eq{8.3}), we consider
\be
a_\mu\to 0,\qquad {a_i a_j a_k \over b}\to h_{ijk},\qquad
{a_i\over b}\to\pm\infty,\qquad A^\mu_\nu\to\hat{A}^\mu_\nu
\label{8.16} \ee that is, essentially the limit \eq{8.8} (to
simplify the presentation, no coordinates of order $b^{2/3}$ are
considered since they lead to the results of \S5). However, it
will become evident that the results obtained are valid for {\em
any} limit of the type \eq{8.15}.

By \eq{8.16} we have
\be
a_i = \b \a_i, \qquad \a_i \a_j \a_k  \to h_{ijk}, \qquad \b :=
b^{1/3} \label{8.17} \ee For arbitrary $f \!\! \in \! \!\A$, $\d
f$ is given by \eq{5.5} and we consider its value in the limit
\eq{8.16}: Expanding \eq{5.5'} and using that $\sum_\mu B^\mu_i =
\sum_\mu A^0_\mu B^\mu_i = 0$, once again we get (\ref{8.10}i),
namely \bez \bar{\pa}_i f = \pa_i f + o(\b) \eez Similarly,
\eq{5.5''} gives
\be
{1 \over 2} \Delta f &=& {1 \over 2 \b} \sum_{j , k} \left(
\sum_\mu \left(\a_j A^j_\mu \right)\left(\a_k A^k_\mu
\right)B^\mu_0 \right) \pa_j \pa_k f \nonumber \\ &+& {1 \over 6}
\sum_{j , k , l} \left( \sum_\mu \left(\a_j A^j_\mu
\right)\left(\a_k A^k_\mu \right)\left(\a_l A^l_\mu \right)
B^\mu_0 \right) \pa_j \pa_k \pa_l f + o(\b) \label{8.18} \ee To
simplify the notation, we put \bez A_\mu(\xi) := \sum_j \a_j
A^j_\mu \xi_j \to \hat{A}_\mu (\xi), \qquad \xi := (\xi_j) \in
\mathbb{R}^N \eez and we consider the functions
\be
\vartheta_2(\xi) = {1 \over 2 \b} \sum_\mu \left( A_\mu (\xi)
\right)^2 B^\mu_0, \qquad \vartheta_3(\xi) = {1 \over 6} \sum_\mu
\left( A_\mu (\xi) \right)^3 B^\mu_0, \label{8.19} \ee related to
the coefficients in \eq{8.18}. We notice that by the Schwartz
inequality \bez \left| A_\mu(\xi) \right| \leq \ \parallel \! \xi
\! \parallel \ \parallel \! A_\mu \! \parallel \ \leq \ \parallel
\! \xi \! \parallel  \max_\mu \{ \parallel \! A_\mu \! \parallel
\}
 =: M \parallel \! \xi \! \parallel \ \to \hat{M} \parallel \! \xi \! \parallel
\eez where $A_\mu=(\a_jA^j_\mu) \in \mathbb{R}^N$ and $\parallel \
\parallel$ is the Euclidean norm in $\mathbb{R}^N$. Therefore,
\be
\left|\vartheta_3(\xi) \right| \leq {M \parallel \! \xi \!
\parallel \over 6} \sum_\mu \left(A_\mu(\xi) \right)^2
\left|B^\mu_0 \right| \ \to {\hat{M} \parallel \! \xi \! \parallel
\over 6} \sum_\mu \left(\hat{A}_\mu(\xi) \right)^2 \hat{B}^\mu_0
\label{8.20} \ee where we used that $B^\mu_0 \to \hat{B}^\mu_0
\geq 0$, eq \eq{5.14}.

On the other hand, if we require that \eq{8.18} is well defined in
the limit \eq{8.17}, then $\sum_\mu \left( A_\mu (\xi) \right)^2
B^\mu_0$ in (\ref{8.19}i) should be at least of order $\b$ (cf.\
\eq{8.12}) and therefore, its limiting value should be zero, i.e.
\bez \sum_\mu \left( \hat{A}_\mu (\xi) \right)^2 \hat{B}^\mu_0 = 0
\eez Therefore, $\vartheta_3(\xi) = 0$ for all $\xi \in
\mathbb{R}^N$, that is, the coefficients of 3rd derivatives of $f$
in \eq{8.18} vanish.

We remark the following:\\ (i) From the form of \eq{8.18},
\eq{8.19}, it is clear that if we require $\d f$ to be
well-defined in the limit \eq{8.17}, then  {\em all} coefficients
of the derivatives of order higher than the second in the
expansion of $\d f$, eq\eq{5.5}, vanish.\\ (ii) The same
conclusion holds if we consider the limit \eq{8.15}, that is,
instead of \eq{8.17}, $a_i = \b^{1/L} \a_i, \, L \geq 3$.\\ (iii)
For the above result, it is {\em essential} that $\hat{B}^\mu_0
\geq 0$. This is a consequence of our interpretation of vector
fields as transition probability distributions. In fact, the
procedure followed above, is a modification of an argument used in
the theory of Markov processes which ensures that higher than the
second moments of such processes vanish (\cite{Gard1985},
\S3.4).\\ (iv) It can be seen from \eq{8.18}, that in the limit
\eq{8.17}, the resulting second order coefficients are no longer
nonnegative-definite (see also \eq{8.14}).

\smallskip
Therefore, summarizing our results in this section, we may say
that a continuous description can be obtained from {\em any}
discrete lattice structure only if the discrete phase space
coordinates are at least of order $\sqrt{t}$, where $t$ is the
time. In this case we get evolution equations at most of the 2nd
order with nonnegative-definite 2nd order coefficient, which is
the most general (linear autonomous) differential generator
conserving probabilities (\cite{TGr1999} \S\S4,5).

\section{Discussion}
\setcounter{equation}{0}

In this section we summarize the basic assumptions introduced in
this paper, results and conceptual insights obtained and we
comment on the limitations of the present approach and on
directions in which it can be further elaborated.

We started with a discrete picture of the extended phase space
$\M$. This corresponds to looking at all fine structural details
of the system and subsequently pass to its ``coarser''
description, which corresponds to the limiting procedures followed
in \S\S4--8, as for instance done in statistical mechanics and
lattice field theory.

On the other hand, we saw in \S3 that a discrete picture allows
for a geometric visualization of the universal DC on $\M$ as a
di-graph, from which special DC on $\M$ can be easily constructed.
In this way, their necessarily noncommutative character discussed
in \S3.1 is interpreted as due to the non-vanishing ``size'' of
the differentials. In fact, these general ideas were illustrated
in \S2. The 1D discrete model studied there, clearly suggested
connections between motion defined on $\M$ endowed with a
noncommutative DC, and a random walk on an oriented square
lattice, one dimension of which is related to time (cf.\
\eq{2.4}). In \S3, vector fields\ $X$ were defined as elements of
the space dual to 1-forms and turn to be 1st order difference
operators on the algebra $\A$ of functions on $\M$, when $\M$ is
an ($N+1$)-dimensional oriented hypercubic lattice. Here, a {\em
crucial} fact is that only vector fields of a {\em special} form
are the generators of automorphisms of $\A$, in contrast to the
usual (commutative) DC, where {\em any} vector field\  $X$ (i.e.\
1st order differential operator) generates (local) automorphisms
of $\A$. That a vector field\ generates automorphisms of $\A$ is
equivalent to the fact that it induces a flow of {\em well
defined} trajectories on $\M$. In the discrete case, these are
paths which connect points of the lattice. This raises the
question of what kind of motion is described by an {\em arbitrary}
vector field Thus, we have been led to a new way of looking at
vector fields, namely, both as {\em generators of evolution of
observables} and as {\em states describing transition
probabilities} for ``infinitesimal'' changes on the lattice.

In this new perspective, probabilistic concepts are introduced in
the dynamics at the ``infinitesimal level'', in contrast to
(classical) statistical mechanics, where microscopic dynamics and
probability distributions are two {\em a priori} quite distinct
concepts. Mathematically speaking, this {\em double role} of
vector fields\ stems from the fact that in the present context,
the concept of a vector field \  is much more general than the
generator of an automorphism of $\A$, as in the case of classical
dynamics based on ordinary DC (cf.\ the discussion at the end of
\S3.2). In fact, this is a {\em general} feature of noncommutative
DC on both discrete and ``continuous'' manifolds, which allows for
the description of evolution equations involving higher than the
1st derivatives {\em and/or} difference operators: By using the
concept of a vector field, a noncommutative version of
differential geometry and tensor calculus can be developed in
close analogy with the ordinary (commutative) case. This is of
potential value in many areas of physics (see e.g.\
\cite{BDM-H1995}, \cite{DM-HV1995}, \cite{DT1996}). We will come
back to this point at the end of this section.

An indication for the consistency of the interpretation of vector
fields\  $X$ as giving the transition probability distribution for
``infinitesimal'' motion along the lattice axes, is provided by
the proof that the associated correlation matrix
$\mathbb{P}=(P^{\mu \nu})$ for the differentials of the lattice
coordinates, vanishes if and only if $X$ generates automorphisms
of $\A$, or equivalently, if trajectories along the lattice are
well defined. This means that there is no possible interference of
motions along {\em different} axes at the {\em same} point (motion
along a specific axis is either impossible, or certain). Moreover,
in this way it became clear that, a function $f$ on $\M$ is an
observable, in the sense that expectation values for its change
represented by $\d f$, are given by $\langle \d f , X \rangle$,
multiplication of observables corresponding to the $\bu$-product
in $\O^1$, eq(\ref{1.3}). On the basis of this and assuming on
physical grounds that time flows with certainty, i.e.\ for every
vector field\  $X$ there is {\em always} a change of some fixed
element, $t$ say, of $\A$, it follows that $\d t$ is proportional
to the (essentially) unique eigenvector of the above mentioned
correlation matrix $(P^{\mu \nu})$ belonging to the zero
eigenvalue. It turns out to be the unit $\rho$ of the commutative
algebra $(\O^1(\A), \bu)$ of 1-forms, $\d t = -b \rho$ (\S\S3.3,
3.4). Using these results, the generator of dynamical evolution,
$X$, is determined by
 $N$ 1-forms $\alpha^i$ vanishing along $X$, eq(\ref{5.**})
and the time evolution of observables is given by
 $-X(f)/b = 0$, eq(\ref{3.33}), in direct analogy with classical dynamics.

In \S\S4, 5, using the probabilistic framework of \S3, we
considered appropriately scaled linear coordinate transformations
on the lattice. We passed to the continuous limit in which $\M$
becomes $\mathbb{R}^{N+1}$ in the same way this is done in the
theory of brownian motion. We showed that the evolution equation
for observables is a 2nd order partial differential equation with
{\em nonnegative-definite} leading coefficient. This is the limit
of the correlation matrix $(P^{\mu \nu})$ of (changes along) the
lattice axes, namely $(\hat{\eta}^{ij})$ in (\ref{5.16}) or
(\ref{5.21}). Each of its elements measures the deviation of the
evolution operator from an ordinary derivation (i.e.\ generator of
deterministic motion) in the corresponding phase space axis. Thus,
noncommutativity of the DC in the {\em continuous} limit (that is,
in a ``coarse-grained'' picture of the system's evolution) is due
to the fact that on a discrete (``fine'') level, motion described
by a vector field, is in general not along well defined
phase-space trajectories. This is an idea also appearing in the
theory of brownian motion and stochastic mechanics, though in a
completely different mathematical and conceptual framework (see
e.g.\ \cite{N1967}, \cite{N1985}, \cite{Blanch1987}). In fact, the
explicit form of  $(\hat{\eta}^{ij})$ in (\ref{5.21}) (cf.
(\ref{5.20}) as well) is consistent with the form of the diffusion
matrix in the kinetic theory of open systems, the theory of
markovian stochastic processes and StC. Moreover, the coefficients
of 1st order derivatives in the evolution equation (\ref{5.17}),
are just the first moments of the coordinate changes with respect
to the transition probability distribution defined by the vector
field\  $X$ (cf.\ (\ref{5.11}), (\ref{5.12})). Therefore, our
evolution equation (\ref{5.17}) is the formal adjoint of a
generalized FP equation (backward equation), justifying our
interpretation of $f \in \A$ as observables and thus giving still
another indication of the consistency of our approach. An
interesting conclusion in this context is that the interpretation
of vector fields\ as transition probability distributions on the
lattice, implies in the {\em continuous limit} that {\em evolution
is forward in time}\/. Thus, if we accept the correspondence $$
\begin{array}{ccc}
\mbox{discrete description} &\longleftrightarrow& \mbox{fine
details of the system's microstructure}\\ \downarrow & &
\downarrow\\ \mbox{continuous description} &\longleftrightarrow&
\mbox{``coarse grained'' (macroscopic) picture}
\end{array}
$$ then, in the present conceptual framework, irreversible
evolution is a characteristic of macroscopic systems.

The continuous limit considered in \S5, at first sight may appear
an arbitrary choice among many possible ones that would lead (via
the expansion of \eq{5.8}) to evolution equations in general
involving derivatives of any order. This issue was studied in \S8
and we have shown that (i) the limiting procedure of \S5 is the
only way to pass  to a continuous description by scaling the
lattice coordinates, {\em without} imposing additional {\em ad
hoc} conditions on the DC, that is, it is the only continuous
limit independent of the general form of the DC one starts with;
(ii) if the probabilistic interpretation of vector fields
introduced in \S3 is employed, then for all types of limiting
procedures, higher than the second order differential equations
are impossible as evolution equations. Thus, (i) and (ii) imply
the unique character of the limit considered in \S5.

In \S\S4, 7 we considered simple applications of our approach:
\smallskip

\noindent (a) The derivation of the 1D diffusion and Smoluchowski
equations for a constant external field, and the 1D FP equation in
velocity space giving the Ornstein-Uhlenbeck process, thus
illustrating the possibility to incorporate random walk models in
the present context.
\smallskip

\noindent (b) The derivation of Kramers' equation in one spatial
dimension, by assuming (i) (newtonian) motion under friction
linear in the velocity and an external field; (ii) trajectories
exist in {\em configuration} space, that is, motion is
deterministic there. In fact, by using the explicit form of the
correlation matrix $\hat{\eta}^{ij}$ under the above assumptions,
we have shown that the diffusion coefficient in our evolution
equation coincides with the diffusion coefficient of Kramers'
equation computed in the context of the theory of brownian motion,
or of kinetic theory.

On the other hand, our result makes explicit the use of assumption
(ii) above, a fact often hidden in the derivations of Kramers'
equation from physically plausible stochastic models. Although
this is also made explicit when Kramers' equation is derived from
Langevin's equation seen as a stochastic differential equation, in
our opinion, the present approach has some advantage: In the
context of StC, it is not easy to relax this assumption (see the
discussion following (\ref{7.2})). In the present context however,
this is possible. Actually, the general form of the corresponding
corrections is evident, namely, a diffusion term in the
configuration space, much in accordance with what seems plausible
on the basis of the (classical and quantum) kinetic theory of open
systems. Finally, it is clear that these results are valid in the
3D case.

The approach elaborated in this paper, clarifies several {\em
conceptual} issues concerning the relevance of noncommutative DC
to (the derivation of) kinetic equations and to StC. In addition,
by giving a definite prescription for the general form of
irreversible evolution equations in terms of vector fields\  $X$
in the discrete framework, it provides an adequate discrete
formalism for deriving such equations in an appropriate continuous
limit. However, it should be supplemented by a ``dynamics'', that
is, a general procedure for {\em choosing} the generator $X$ in
particular cases. In sections \S 5 and 7, we have used Newton's
equations written as {\em 1-form relations} (see (\ref{5.*}) and
the derivation of (\ref{5.12}), (\ref{7.12})). As briefly
discussed at the end of \S7, a more systematic approach is to
develop symplectic mechanics in the present context and require
$X$ to be hamiltonian, in analogy with classical mechanics. In
fact, from a {\em mathematical} point of view, this approach has
been followed successfully in the {\em continuous} regime for the
second order DC defined by (\ref{5.18}), giving the promising
result that hamiltonian evolution equations have the form of
generalized FP equations that appear in kinetic theory
(\cite{DT1996}). In another paper, the present approach will be
elaborated in this direction.

It is also possible (and physically desirable) to consider the
extension of the present formalism when the transformation from
the lattice coordinates $(u^\mu)$ to the scaled coordinates
$(t,x^i)$, eq(\ref{5.1}), is not linear. Many of the present
results are expected to be still valid.

Finally, the general formalism and the conceptual framework in
this paper, may be extended from the case of the oriented
hypercubic lattice, to more general structures induced by
appropriate DC on a discrete manifold. More precisely, it is clear
that throughout this work, {\em differentials are not }
necessarily {\em infinitesimal quantities} in the usual {\em
geometric} sense of ordinary DC. Rather than that, if seen in the
suggestive representation of a DC as a di-graph, they express {\em
interrelations} between points of the discrete manifold, so that
all points related to a given one, should be considered as being
{\em neighboring} to it (see (\ref{3.6}), its interpretation in
\S3.1 and the differentials for the hypercubic lattice,
eq(\ref{3.21'})). This idea, that a DC is based on a concept of
``relational'', rather than geometric type of ``proximity'' may be
further elaborated. In this way, it may become possible to develop
a general mathematical and conceptual framework for describing
physical systems based on the {\em interrelations} (interactions)
among its different parts, rather than on their relative {\em
geometric} position. This is virtually relevant in situations
where collective effects are significant, or even dominant (e.g.\
systems with long-range interactions), for which traditional
approaches often do not work beyond the lowest approximation. We
will come back to this point in another work.

\section*{Appendix A}
\renewcommand{\theequation}{A.\arabic{equation}}
\setcounter{equation}{0}

{\bf 1.}  Here we prove the following proposition referred to in
\S3.2.
\medskip

\noindent {\bf Proposition} {\em A 1-1 and onto mapping $\Phi: \M
\mapsto \M$ induces an automorphism $\phi: \A \mapsto \A$ that
maps the basis $\{ e_i \, , \; i \in \M \}$ of $\A$ onto itself.
The converse is also true.}
\medskip

\noindent {\bf Proof:} ``$\Rightarrow$'': We define
\be
\phi \!: \A \mapsto \A \, : \quad \phi(f)(i) = f(\Phi(i))
\label{A1} \ee Clearly $\phi$ is an endomorphism and \bez
\phi(e_i)(j) = e_i(\Phi(j)) = \delta_{i, \Phi(j)} =
\delta_{\Phi^{-1}(i),j}= e_{\Phi^{-1}(i)}(j) \eez hence
\be
\phi(e_i) = e_{\Phi^{-1}(i)} \label{A2} \ee which shows that
$\phi$ is 1--1 and onto.

``$\Leftarrow$'': Conversely, let $\phi$ be an automorphism of
$\A$. From (\ref{3.3}i) we get $\phi(e_i) (\phi(e_j)-\delta_{ij})
= 0$. If $\phi(e_i)=\sum_k f_{ik} e_k$, the above equation implies
that $f_{ik} = 0$ or $1 \, , \; \forall \, i, k$. By (\ref{3.3}ii)
$\sum_{i,k} f_{ik} e_k = \sum_k e_k$ hence $\sum_i f_{ik}=1$ where
we have used that $\phi(1)=1$ and $\{ e_i \}$ is linearly
independent. Consequently, in each column $k$ of $f_{ik}$ there
exists exactly one element, on the $i_k$-th row, say, which is
nonzero and therefore, necessarily equal to 1. Now suppose that
for $k \neq l$, the corresponding nonzero elements are on the same
$i$-th row. Then, these columns are identical which is impossible,
since $\phi$ is 1-1 and onto. Therefore, in each row $i$ of
$f_{ik}$, exactly one $k$ is nonzero and equal to 1. Hence,
$\phi(e_i)=e_k$ for some $k$ depending on $i$. Now,
$\phi^{-1}(e_i)=e_{\Phi(i)}$ defines $\Phi: \M \longmapsto \M$
which is 1-1 and onto and satisfies (\ref{A1}). \hfill QED
\bigskip

\noindent {\bf 2.} By considering the case of classical dynamics,
we explain below the choice of the sign in (\ref{3.32}) and
comment on the form of (\ref{5.**}).

Let the  equations of motion of an $N$-dimensional dynamical
system be \bez {\d x^i \over \d s} = R^i(x) \eez The time
evolution of observables $f$ and states $\sigma$, is given by
semigroups of Koopman and Perron-Frobenius operators respectively,
whose infinitesimal generators lead to \bez {\pa f \over \pa s} =
R^i(x){\pa f \over \pa x^i}\, , \qquad {\pa \sigma \over \pa s} =
-{\pa  \over \pa x^i}\left(R^i(x)\sigma \right) \eez with the
summation convention used here. The operators on the r.h.s. are
formal adjoints to each other (cf. e.g.\ \cite{LM1985} ch.7).

Then, the prescription for {\em extended dynamics} is
\be
\begin{array}{lcccc}
\mbox{equations of motion:}&\hspace{1cm} &\displaystyle{{\d x^i
\over \d s}} = R^i(x) \, ,&
  & \displaystyle{ {\d t \over \d s}}=1\\[0.3cm]
\mbox{evolution of states:}& &\displaystyle{\left( {\pa \over \pa
t} + {\pa \over \pa x^i} R^i(x) \cdot\right)} \sigma & =:&
X_S(\sigma)=0\\[0.4cm] \mbox{evolution of observables:}&
&\displaystyle{\left( -{\pa \over \pa t} + R^i(x){\pa \over \pa
x^i} \right)}f& =:&
 X_O(f)=0
\end{array}\hspace{1.7cm} \label{A3}
\ee Therefore, the system of characteristic equations of the
evolution equation for {\em observables}, is given by the vector
field with components $(X_O(t), X_O(x^i)) = (-1, R^i)$, i.e.\
evolution is obtained by integrating the equations of motion
``backwards in time''.\footnote{cf.\ the theory of stochastic
processes and stochastic differential equations (e.g.\
\cite{Arn1973}, \S2.6), where, for observables (i.e.\ bounded
functions of a stochastic process): Generator of the evolution of
observables = ({\em backward} generator of evolution of the
probability distribution)=
 $-$(adjoint of {\em forward} generator of evolution of the
probability distribution).}
 This explains the $(-)$ sign in (\ref{3.32}) and the form of
(\ref{5.*}). In fact, to the system of characteristic equations of
the observables' evolution equation (\ref{A3}iii) are associated
the 1-forms
\be
\alpha^i = \d x^i + R^i \d t \qquad \quad i=1,2, \ldots, N
\label{A4} \ee vanishing along the integral curves of the vector
field $X_O$. This is equivalent to
\be
\langle \alpha^i , X_O \rangle = 0 \qquad \qquad i=1,2, \ldots, N
\label{A4'} \ee $\langle \ ,  \rangle$ denoting the contraction of
1-forms with vector fields in the ordinary DC. In the conceptual
framework of the present paper, this is given by (\ref{5.*}),
(\ref{5.**}). It is a completely general formulation that includes
as special cases (in an obvious notation)\\ Newton's equations
($N=2M$) \bez \beta^r= \d q^r + v^r \d t \, , \qquad  \gamma^r =
\d v^r + f^r \d t \eez or Hamilton's equations \bez \beta^r= \d
q^r + {\pa H \over \pa p_r}  \d t \, , \qquad \gamma_r = \d p_r -
{\pa H \over \pa q^r} \d t \eez To connect this with the usual
formulation of symplectic mechanics, we notice that, by
introducing the symplectic matrix \bez (J^{ij}) = (J_{ij})=\left(
\matrix{ 0 & I \cr -I & 0} \right) \eez ($I$ being the $M \times
M$ identity matrix), the 1-forms above can be rewritten in a form
identical to (\ref{5.*}), namely \bez \alpha^i = \d x^i + \sum_j
J^{ij} {\pa H \over \pa x^j} \d t \quad \Longleftrightarrow \quad
\alpha_i = \sum_j \ J_{ij} \d x^j -  {\pa H \over \pa x^i} \d t
\eez ($i=1,2, \ldots , 2M$) so that the $\alpha_i$ can be given in
terms of the symplectic form \bez \omega := {1 \over 2} \sum_{i,j}
J_{ij} \d x^i \wedge \d x^j - \d H \wedge \d t, \qquad a_i =
i_{\pa / \pa x^i} \omega \eez where $i$ is the interior product
operator for ordinary differential forms.

\noindent {\bf 3.} Some of the results in \S3 can be seen in the
light of the discussion in the previous subsection: By
(\ref{3.24}), $X(u^\mu)=P^\mu$ is the defining relation for the
components of an arbitrary vector field $X$. The approach in this
paper leads to concrete results in particular cases, once $X$ is
chosen appropriately in each case, say $\langle \d u^\mu , X
\rangle =: P^\mu = \tilde{R}^\mu$, as it was done for instance in
\S4. This fact can be recast  in a more suggestive form, namely
\be
\langle \tilde{\alpha}^\mu, X \rangle =0, \qquad \mbox{where}
\qquad \tilde{\alpha}^\mu := \d u^\mu - \tilde{R}^\mu \rho
\label{A5} \ee that is, the $(N+1)$ 1-forms $\tilde{\alpha}^\mu$
vanish along $X$. Notice that, owing to $\sum_\mu P^\mu =1$, we
must impose the condition $\sum_\mu \tilde{R}^\mu =1$, hence
$\sum_\mu \tilde{\alpha}^\mu =0$, that is, the
$\tilde{\alpha}^\mu$ are linearly dependent. Moreover, by
\eq{3.31'}, the correlation matrix of the lattice coordinates is
$P^{\mu \nu} = \langle  \tilde{\alpha}^\mu  \bu
\tilde{\alpha}^\nu, X \rangle$ in close analogy to \eq{5.22'}. In
fact, (\ref{A5}ii) is just \eq{5.*} written in the lattice
coordinates $(u^\mu)$. To see this, we transform \eq{A5} to $(t,
x^i)$ coordinates. Since $P^\mu=\tilde{R}^\mu$, the $
\tilde{\alpha}^\mu$ become \bez \alpha^0 &=& \sum_\mu (-b) A^0_\mu
\tilde{\alpha}^\mu = -b \sum_\mu  \tilde{\alpha}^\mu = 0  \\
\alpha^i &=& \sum_\mu a_i A^i_\mu  \tilde{\alpha}^\mu = \sum_\mu
a_i A^i_\mu  \d u^\mu - \sum_\mu a_i A^i_\mu \tilde{R}^\mu \rho =
\d x^i + {a_i \over b} (A \tilde{R})^i \d t \eez By setting
$(a_i/b)(A\tilde{R})^i = R^i$, the second equation becomes
\eq{5.*}, hence (\ref{A5}i) is transformed to \eq{5.**}.

\section*{APPENDIX B}
\renewcommand{\theequation}{B.\arabic{equation}}
\setcounter{equation}{0}

As mentioned in \S5 remark (5), in specific cases, instead of
expanding the evolution equation (\ref{5.8}) as in (\ref{5.9}), it
is possible to write it explicitly in terms of appropriate
difference operators which in the continuous limit gives the
corresponding form of (\ref{5.17}). As an illustration, we present
here an $N$-dimensional example which generalizes that of \S2. We
employ the definitions and notations of \S\S3, 5.

The transformation (\ref{5.1}) is chosen as follows
($i=1,2,\ldots,N$)
\be
t=-b\sum_\mu u^\mu \, , \qquad
x^i=a_i(u^0+\cdots+u^{i-1}-u^i+u^{i+1}+\cdots+u^N).  \label{B1}
\ee so that its inverse (\ref{5.4})and the commutation relations
(\ref{5.6}) become
\be
u^0={n-2\over2}\,{t\over b}+{1\over2}\sum_{i=1}^N\, {x^i\over
a_i},\qquad \qquad u^i=-{1\over 2}\left({t\over b}+{x^i\over
a_i}\right), \label{B2} \ee
\be
\begin{array}{c}
\d t\bu\d t =-b\, \d t , \qquad \d t\bu\d x^i =-b\, \d x^i, \qquad
\d x^i\bu\d x^i=- \displaystyle{{a_i^2\over b}}\, \d t, \\[0.4cm]
\d x^i\bu\d x^j = -\displaystyle{{a_i a_j\over b}} \, \d t + a_j\,
\d x^i+ a_i\, \d x^j\
 \end{array} \label{B3}
\ee In the limit (\ref{5.3}),  eq(\ref{5.18}) takes the form
\be
\d t\bu\d t=0,\qquad\d t\bu\d x^i=0,\qquad \d x^i\bu\d x^j=-
h_{ij} \, \d t , \label{B4} \ee

For the differential of a function of $u^\mu$ \bez \d f =\sum_\mu
[f(u^0,\ldots,u^{\mu-1},u^\mu+1,u^{\mu+1},\ldots,u^n)-
   f(u^0,\ldots,u^n)] \, \d u^\mu.
\eez As a consequence eq(\ref{5.5}) takes the form
\be
\d f & = & {1\over 2b}\left[-\sum_{i=1}^N f(t-b,x+a,x^i-a_i,x+a)+
(n-2) f(t+b,x+a)+2 f(t,x)\right] \, \d t\nonumber\\ &&+\sum_i
(\bar{\pa}_i f)(t-b,x+a,x^i,x+a) \, \d x^i, \label{B5} \ee with
\bez (\bar{\pa}_i f)(\ldots,x^i,\ldots):={1\over 2a_i}
[f(\ldots,x^i+a_i,\ldots)-f(\ldots,x^i-a_i,\ldots)] \eez After a
lengthy calculation we obtain
\be
\d f & = &[(\pa_{-t}f)(t,x)-\sum_{i=1}^N{a_i^2\over 2b}\,
(\Delta_i f)(t-b,x+a,x^i,x+a) \nonumber\\ &&+\sum_{1\leq i<j\leq
N}{a_ia_j\over b}\,(\pa_{+i}\pa_{+j}f) (t-b,x+a,x^i,x,x^j,x+a)] \,
\d t \nonumber\\ &&+\sum_i (\bar{\pa}_i f)(t-b,x+a,x^i,x+a) \, \d
x^i, \label{B6} \ee where \bez (\pa_{-t} f)(t,x):={1\over
b}[f(t,x)-f(t-b,x)] \eez \bez (\Delta_i f)(\ldots,x^i,\ldots):=
{1\over a_i^2}[f(\ldots,x^i+a_i,\ldots) +f(\ldots,x^i-a_i,\ldots)
-2 f(\ldots,x^i,\ldots)]. \eez (cf. (\ref{2.7})-(\ref{2.10})).
Evidently, in the limit (\ref{5.3}), eq(\ref{B6}) gives
\be
\lefteqn{\d f =  [(\pa_t f)(t,x)-\sum_{i=1}^N h_{ii} \, (\pa^2_i
f)(t,x) +} & & \hspace{8cm} \nonumber\\ & &
\hspace{3cm}\sum_{1\leq i<j\leq N} h_{ij} \,(\pa_i\pa_j f)(t,x)]\,
\d t +\sum_i (\pa_i f)(t,x) \, \d x^i \label{B7} \ee

By (\ref{3.24}), a vector field\  in this calculus has the form
\bez (Xf)(u) = \sum_{\mu=0}^N P^\mu(\pa_{+\mu}f)(u)=
\sum_{\mu=0}^N P^\mu(f(u,u^\mu+1,u)-f(u)), \eez or, by (\ref{B1})
\be
(X f)(t,x) & = &
X^t(t,x)\,[(\pa_{-t}f)(t,x)-\sum_{i=1}^N{a_i^2\over 2b}\,
(\Delta_i f)(t-b,x+a,x^i,x+a) \nonumber\\ &&+\sum_{1\leq i<j\leq
N}{a_ia_j\over b}\,(\pa_{+i}\pa_{+j}f)
(t-b,x+a,x^i,x,x^j,x+a)]\nonumber\\ &&+\sum_i
X^i(t,x)\,(\bar{\pa}_i f)(t-b,x+a,x^i,x+a),\label{B8} \ee
\be
X^t=-b(P^0+\cdots+P^n),\qquad
X^i=a_i(P^0+\cdots+P^{i-1}-P^i+P^{i+1}+\cdots+P^n) \label{B9} \ee
Then by (\ref{3.18}), $X$ defines a flow on $\cal{M}$, if and only
if
\be
P^\mu\,P^\nu = \delta ^{\mu\nu}\,P^\mu \ee and has $N+1$ solutions
with $P^\mu=1,\,P^\nu=0$, if $\nu\neq\mu$, for $\mu=0,1,\ldots,N$.
Using (\ref{B9}) we readily get that in $(t,x^i)$ coordinates, the
evolution generator for observables, eq(\ref{3.33}), has
components $(1, \pm a_i/b)$, thus obtaining the $N$-dimensional
generalization of the ``random-walk model'' of \S2 (see
eq(\ref{2.17})).

\vspace*{0.5cm} {\bf Acknowledgement}: The authors would like to
thank F. M\"{u}ller-Hoissen for his critical remarks on the
original draft of the paper. C.T. was partially supported by the
University of the Aegean, under grant EPEAEK/397 and the paper was
completed while he was at the University of the Aegean, on leave
from the University of Crete. He would also like to thank all the
members of the Department of Mathematics of the University of the
Aegean for their hospitality.

\end{document}